\documentclass[prb,aps,twocolumn,showpacs,preprintnumbers,nofootinbib]{revtex4}
\usepackage[dvips]{graphicx}
\usepackage{latexsym}
\usepackage{amsmath}
\begin{document}
\renewcommand{\(}{\left (}
\renewcommand{\)}{\right )}
\renewcommand{\i}{\mathrm{i}}
\newcommand{\vect}[1]{\boldsymbol{#1}}
\def\sgn{\text{sgn}}
\def\vev#1{\langle{#1}\rangle}
\newcommand{\bea}{\begin{eqnarray}}
\newcommand{\eea}{\end{eqnarray}}

\title{A controlled expansion for certain non-Fermi liquid metals}

\preprint{MIT-CTP/4127}

\author{David F. Mross, John McGreevy, Hong Liu, and T. Senthil}
\affiliation{ Department of Physics, Massachusetts Institute of
Technology, Cambridge, Massachusetts 02139}

\date{\today}
\begin{abstract}
The destruction of Fermi liquid behavior when a gapless Fermi surface is coupled to a fluctuating gapless boson field is studied theoretically. This problem arises in a number of different contexts in quantum many body physics. Examples include fermions coupled to a fluctuating transverse gauge field pertinent to quantum spin liquid Mott insulators, and quantum critical metals near a Pomeranchuk transition. We develop a new controlled  theoretical approach to determining the low energy physics. Our approach relies on combining an expansion in the inverse number ($N$) of fermion species with a further expansion in the parameter $\epsilon = z_b -2$ where $z_b$ is the dynamical critical exponent of the boson field. We show how this limit allows a systematic calculation of the universal low energy physics of these problems.
The method is illustrated by studying spinon fermi surface spin liquids, and a quantum critical metal at a second order electronic nematic phase transition. We calculate the low energy single particle spectra, and various interesting two particle correlation functions. In some cases deviations from the popular Random Phase Approximation results are found. Some of the same universal singularities  are also calculated to leading non-vanishing order using a perturbative renormalization group calculation at small $N$ extending previous results of Nayak and Wilczek. Implications for quantum spin liquids, and for Pomeranchuk transitions are discussed. For quantum critical metals at a nematic transition we show that the tunneling density of states has a power law suppression at low energies.

\end{abstract}
\newcommand{\be}{\begin{equation}}
\newcommand{\ee}{\end{equation}}
\maketitle

\section{Introduction}
This paper is concerned with the destruction of Fermi liquid behavior in two dimensional systems where a gapless Fermi surface is coupled to a fluctuating gapless boson field. This problem arises in a number of different contexts in quantum many body physics. A well known example is where the fermions are coupled to a gapless transverse gauge boson. This describes the low energy effective theory of certain quantum spin liquid phases\cite{lesik,leesq}, the theory of the half-filled Landau level\cite{hlr}, and various non-fermi liquid metallic phases\cite{lee89,vrtxmtl,holonm,acl,dbl}. A different and equally well known example is as a description of quantum critical metals at a `Pomeranchuk' instability. The classic example is the Stoner transition  associated with the onset of ferromagnetism in a metal. In recent years attention has focused on a different example of a Pomeranchuk transition: that associated with the onset of electronic nematic order\cite{kivfr,metzner,kee,lawler,chubukov,rosch,chubukovfl} from a Fermi liquid metal. Here electronic nematic order means a phase where the lattice point group symmetry but not translation symmetry is broken. Such order has been observed with increasing frequency in a number of different correlated metals\cite{andyM,ando,keimer,kohsaka,taill,nemrev} giving rise to an interest in the associated quantum phase transition. At such a quantum phase transition the nematic order parameter is described as a gapless fluctuating Bose field, and its coupling to the gapless Fermi surface destroys Fermi liquid behavior\cite{metzner,chubukov,lawler}.

The purpose of this paper is to formulate a new controlled theoretical approach to this class of problem, the need for which has been emphasized recently\cite{ssl,maxsubir}. The low energy physics of the resulting non-fermi liquid metal is characterized by universal scale invariant behavior. Our approach provides a systematic method of calculating the exponents and other universal properties associated with this scale invariant behavior. We illustrate this by studying many physical properties of the gauge field model and of the nematic quantum critical metal in detail.

Quite generally the low energy physics of problems of this sort is conveniently described by restricting attention to fermionic modes in the immediate vicinity of the Fermi surface, and the long wavelength, low frequency modes of the fluctuating boson field.  The model is
described by the Euclidean action
\begin{eqnarray}
\label{themodel}
S & = & S_f + S_{int} + S_a \\
S_f & = & \int_{\vec k, \omega} \bar{f}_{k\alpha}\left(- i \omega - \mu_f + \epsilon_{\vec k}\right)f_{k\alpha}  \\
S_{int} & = & \int_{\vec k, \omega} a (\vec k,\omega) O(-\vec k, -\omega) \\
S_a & = & \int_{\vec k, \omega}\frac{1}{e^2}k^2 |a(k,\omega)|^2
\end{eqnarray}
Here $f_{k\alpha}$, $\alpha = 1,....N$, is a fermion field with $N$ possible flavors and $a$ is the boson field. In the gauge model $a$ is the transverse component of a $U(1)$ gauge field, and $O(\vec x, \tau)$ is the transverse component of the current density of fermions. At a nematic quantum critical point $a$ will be taken to be the nematic order parameter field, and $O(\vec k, \omega)$ is the fermion bilinear with the same symmetry. For instance, on a two dimensional square lattice with lattice constant $\ell$ a uniform nematic order parameter couples to
$N^{-1/2} \sum_k \left(\cos(k_x \ell) - \cos(k_y \ell) \right) \bar{f}_{k\alpha} f_{k\alpha}$.

Much prior work of course exists on this problem.
In a number of early papers\cite{holstein,reizer,lee89,leenag,hlr} the problem  was analysed in a Random Phase Approximation (RPA) and various related approaches. This showed that
the fermions and gauge bosons stay strongly coupled in the low energy limit. In the RPA, the boson propagator is overdamped due to Landau damping by the gapless Fermi surface. The fermion self energy has a power law non-fermi liquid frequency dependence. Further the long wavelength density response function retains its Fermi liquid form.

In the gauge field problem, some of the RPA results were further substantiated\cite{ybkim} through a quantum Boltzmann approach which considered the fate of various possible shape fluctuations of the Fermi surface. Smooth shape deformations of the Fermi surface (which determine long wavelength density response and the gauge propagator) were shown to retain Fermi liquid behavior while ``rough" deformations have the potential to be non-fermi liquid like. The latter determine the behavior of the single fermion Green's function and the structure of the $2K_f$ singularities ({\em i.e} at wavevectors connecting antipodal tangential portions of the Fermi surface) in response functions. These main results were further supported in  comprehensive diagrammatic analyses\cite{aliomil} of the model which suggested that the leading RPA answers for many quantities were in fact exact in the low energy limit. In particular the structure of the gauge field propagator, the fermion self energy, and the long wavelength density response were argued to have the same form as the RPA result. Similar diagrammatic analyses with the same conclusions have also been reached\cite{chubukov} for the nematic quantum critical point. In the gauge field case the $2K_f$ singularities in the density response function were argued to have specific non-fermi liquid like power law forms\cite{aliomil}.

Is there a controlled limit in which the reliability of these results may be assessed? One attempted approach\cite{polchinski,aliomil} is to take the limit of $N$ (the number of fermion species) large, and expand in powers of $1/N$. In the early work it was argued\cite{polchinski} that at low energies in the large-$N$ limit only patches of the Fermi surface with parallel normals are strongly coupled to each other. Any such patch couples strongly to a boson whose momentum is perpendicular to the normal to the Fermi surface. The low energy physics is therefore correctly described by focusing attention on patches with parallel normals.

In some remarkable recent work, Sung-Sik Lee\cite{ssl} reexamined the model of $N$ fermion species coupled to a $U(1)$ gauge field in the large-$N$ limit.
He showed that even at large-$N$ the theory remains strongly coupled, and that its solution requires non-trivial summation of an infinite number of Feynman diagrams. When only a single patch of the Fermi surface is considered, a book-keeping device was introduced to  show that the $1/N$ expansion could be organized in terms of the genus of the surface in which the Feynman diagrams were drawn. Based on this the general validity of the physical picture built up by RPA and the other earlier analyses for small $N$ has been questioned.

Even more recently in another very interesting paper Metlitski and Sachdev\cite{maxsubir} studied the fate of the theory with both a Fermi surface patch and its antipodal partner included. This is believed to be fully sufficient to correctly describe the asymptotic low energy physics of the system. They found a number of further difficulties with the large-$N$ expansion. Specifically higher loop corrections  for the gauge propagator involved higher powers of $N$ than the leading order one loop RPA result. This unpleasant finding led them to question the existence of a well-defined large-$N$ limit to control the theory. These authors also showed that in a perturbative loop expansion the self energy acquires singular momentum dependence at three loop order. However the loop expansion has no apparent control parameter.

In light of these results it becomes important to search for alternate reliable methods to judge the validity of RPA and other diagrammatic approaches to the problem. In this paper we introduce a new controlled expansion to determine the low energy physics of this model. We consider a family of models
where the `bare' boson action is modified to
\begin{equation}
S_a  =  \int_{\vec k, \omega} \frac{|\vec k|^{z_b - 1}}{e^2} |a(\vec k, \omega)|^2
\end{equation}
The number $z_b$ (the boson ``dynamical critical exponent") equals 3 in the original model in Eqn.~\ref{themodel}. The case $z_b = 2$ arises in the theory of the half-filled Landau level with long range $1/r$ Coulomb interactions between the electrons\cite{hlr}, and in the theory of the bandwidth controlled Mott transition of the half-filled Hubbard model developed in Ref.~\onlinecite{mottcrit}. We show that the large-$N$ expansion can be controlled in the limit of small $\epsilon = z_b - 2$. Specifically we show that the limit $N \rightarrow  \infty, \epsilon = z_b - 2 \rightarrow 0$ such that $\epsilon N$ is finite leads to reliable answers for the low energy behavior of the system. We demonstrate that the RPA answers for the fermion and boson propagators are indeed exact in this limit. A systematic expansion in powers of $1/N$ is possible for small $\epsilon$.  Deviations from RPA  emerge at higher orders in the $1/N$ expansion.  Furthermore differences between the gauge model and the nematic critical point also appear. At order $1/N^2$, we find a singular momentum dependent correction to the fermion self energy
- however, in the gauge model, this
singularity is subdominant to the leading order
momentum dependence, so that the fermion self energy retains its RPA form, at least to this order.  Further we calculate the exponent characterizing $2K_f$ particle-hole singularities, and show in the gauge model that (depending on the value of $\epsilon N$), they may be enhanced compared to a Fermi liquid.
For a quantum critical metal at a nematic transition the fermion propagator again retains its RPA form at leading order but at $o(1/N^2)$ acquires a singular correction to the self energy that dominates over the RPA form. This modification from the RPA is in accord with the calculation of Ref.~\onlinecite{maxsubir} but is now performed in a controlled expansion. A further difference with the gauge field problem is in the structure of the $2K_f$ singularities. We present calculations and physical arguments that show that the $2K_f$ singularities are weakened at the nematic quantum critical point compared with a Fermi liquid.

A crucial physical ingredient that determines the low energy physics is the nature of the `Amperean' interaction between two fermions that is mediated by the boson field. The term `amperean' is appropriate for the gauge field case where the interaction is between fermion currents but we will use it to describe the nematic transition as well. In the gauge field case, the currents of a particle in one patch of the Fermi surface is parallel to that of a hole in the antipodal patch. By Ampere's law the gauge mediated interaction between such a $``2K_f"$ particle-hole pair is attractive. In contrast the currents of two particles with one from either patch are antiparallel and the gauge mediated interaction is repulsive in the particle-particle (Cooper) channel. The situation is reversed in the nematic model. We show how this difference between the particle-hole and Cooper channel interactions plays an important role in many aspects of the low energy physics.

Previously Nayak and Wilczek\cite{chetan}  studied the low energy physics of the gauge field model for small $\epsilon$, and finite $N$ using  perturbative renormalization group (RG) methods.
The one loop beta function for the coupling constant $e$ takes the form
\begin{equation}
\beta(e^2) = \frac{\epsilon}{2} e^2 - \frac{c}{N}e^4
\end{equation}
where $c$ is a positive constant. This leads to a perturbatively accessible non-fermi liquid fixed point for $N\epsilon$ small and positive. Some properties of this fixed point were calculated in Ref.~\onlinecite{chetan} and shown to be consistent with the RPA analysis.
Calculations with this one loop beta function are able to provide answers for the scaling exponents to order $\epsilon$ for any $N$. In contrast our approach of directly solving the theory at large-$N$ enables us to extract exponents that are high order or even non-perturbative in $\epsilon$.  Thus for instance the deviations from RPA (subdominant in the gauge field model) discussed above are expected to appear in the $\epsilon$ expansion only at order $\epsilon^3$, and so do not show up in the $o(\epsilon)$ calculations. In the regime where they overlap ({\em i.e} to order $\epsilon$ at fixed large-$N$), we will show that the exponent values for many properties calculated within our approach agree with those obtained from the perturbative RG. As it does not seem to be available in the literature, we calculate the $2K_f$ exponent at this small $\epsilon$, finite $N$ fixed point.  We show that the exponent has an interesting non-analytic dependence on $\epsilon$ for small $\epsilon$ which leads in the gauge model to an enhancement of the $2K_f$ singularities compared with a Fermi liquid in this limit. In the nematic model we find a suppression of the $2K_f$ correlations as expected on general physical grounds that we also discuss.

An important feature of the low energy physics is  that the Fermi surface is preserved and is sharp even though the Landau quasiparticle is destroyed. Further at low energies and for momenta close to the Fermi surface the fermionic spectrum is scale invariant. This was already implied by the RPA results, and survives in our treatement.  A  similar picture was also argued\cite{critfs} to describe continuous phase transitions where an entire Fermi surface disappears (such as a continuous Mott transition). Following Ref \onlinecite{critfs} we will refer to this as a critical Fermi surface. Some (though not all) aspects of critical fermi surfaces associated with Mott-like transitions may be expected to be shared with the Pomeranchuk transitions discussed in this paper.
It is therefore useful to consider these results in terms of a general scaling form expected for fermions with a `critical Fermi surface'. We write for the fermion Green's function
\begin{equation}
\label{frmscl}
G(\vec K, \omega) \sim \frac{c_0}{|\omega|^{\frac{\alpha}{z}}} g_0\left(\frac{c_1 \omega}{k_{\|}^z}\right)
\end{equation}
Here $k_{\|}$ is the deviation of the momentum from the Fermi surface. Note that the $z$ that enters this scaling equation is the `fermionic' dynamical critical exponent. For the problems studied in this paper the RPA gives $z = \frac{z_b}{2}$ and $\alpha = 1$. The latter is a result of the absence of any singular momentum dependence in the self energy in RPA. Our results may be viewed as a calculation of $\alpha$ and $z$ within a systematic expansion. In the gauge model to $o(1/N^2)$ these exponents do not change as far as the leading singular structure is concerned. For the nematic critical point we find that $\alpha = 1- \eta_f$ with $\eta_f$ positive.

Following the general discussion in Ref.~\onlinecite{critfs}, we show that the difference from the RPA result has direct and measurable consequences for the electron single particle tunneling density of states $N(E)$ (where $E$ is measured from the chemical potential) at the nematic quantum critical point. Within the RPA the tunneling density of states is a constant at the Fermi level. However beyond RPA there is a power law suppression of $N(\omega)$:
\begin{equation}
N(\omega) \sim |\omega|^{\frac{\eta_f}{z}}
\end{equation}
The exponent $\eta_f$ is calculated in Section \ref{sec:twopatch}. Extrapolation of the leading order results to $z_b = 3, N= 2$ gives the estimate $\eta_f \approx 0.3$.

What about the fate of the large-$N$ limit when $z_b$ is not close to two? We suggest that recent calculations of Ref.~\onlinecite{maxsubir} should be interpreted as an instability towards translational (and possibly other) symmetry breaking in this limit. In the nematic context this means that there is no direct second order quantum phase transition associated with nematic ordering in two dimensions if $N$ is sufficiently large. Instead the transition is preempted by the appearance of density wave and possibly other orders. So if for the physical case $N = 2$ there is a direct nematic transition, then it is not usefully accessed by the large-$N$ expansion. Our approach of combining the large-$N$ with a small $z_b - 2$ or a direct small $z_b - 2$ perturbative RG then become the only available methods to theoretically access such a quantum critical point directly in two dimensions. Similar phenomena also happen in the gauge field model - in the large-$N$ limit we propose that the uniform state is unstable to translation symmetry breaking.

\begin{figure}[t]
\includegraphics[width=80mm]{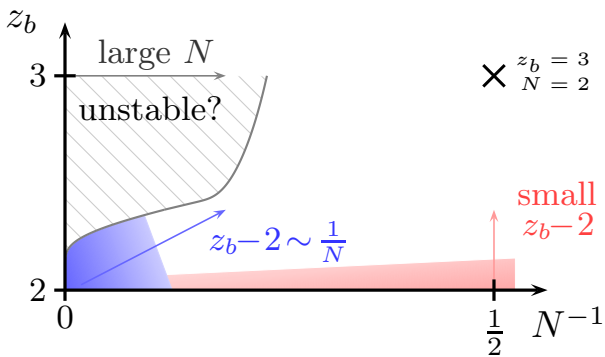}
\caption{(color online) Our suggested phase diagram.
Above the indicated curve, the putative critical theory is likely
preempted by some other broken-symmetry state.
The behavior of the proposed phase boundary at small $N$ is
one possible extrapolation.}
 \label{favor1}
\end{figure}
\begin{figure}[t]
\includegraphics[width=80mm]{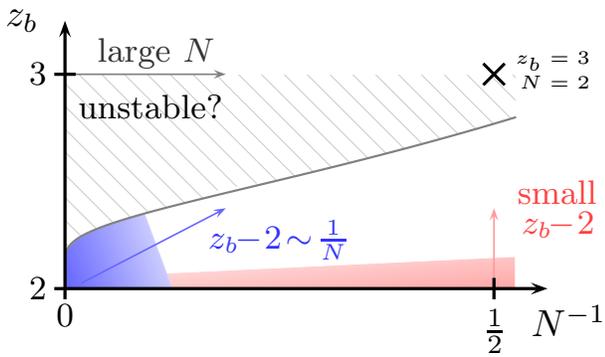}
\caption{(color online) An alternate possible phase diagram.
Here the interesting point $z_b=3, N=2$ is in the unstable regime;
in this case, it would be best accessed starting from the
correct mean field theory for the new broken-symmetry state.
}
 \label{favor2}
\end{figure}

It is instructive to consider the behavior of the model in a two dimensional plane spanned by $z_b$ and $1/N$. We show our proposed `phase diagram' in Figs.~\ref{favor1} and \ref{favor2}. It is clear that the approach developed in this paper is ideally suited to describing the gauge model or the nematic quantum critical point for $N = 2$ if it is not part of the unstable region, {\em i.e} if
Fig.~\ref{favor1} is realized. If on the other hand $z_b = 3, N = 2$ belongs to the unstable region as depicted in Fig.~\ref{favor2}, we may still hope that the approach in this paper is useful in describing the physics at temperatures above the onset of the instability.

The rest of the paper is organized as follows. In Section.~\ref{sec:prelim}, we begin with some preliminaries and briefly discuss the patch construction for the Fermi surface that is used in the rest of the paper. Some subtle but important aspects of the patch construction are relegated to Appendix~\ref{app:anomalies}. Then in Section~\ref{sec:scaling} we warm up by studying the theory of just one patch and ignoring any coupling with the other antipodal patch, and show how our expansion provides a controlled answer in this simplified problem. We then study the full two patch theory in Section~\ref{sec:twopatch} and determine the singular structure of the boson and fermion propagators. In Section~\ref{twokf} we present a calculation of the exponent characterizing $2K_f$ singularities within our approach. Next in Section~\ref{pertRG} we explore the connections with the perturbative RG calculations of Ref.~\onlinecite{chetan} and extend their results to $2K_f$ singularities. In Section~\ref{sec:physics} we discuss simple physical interpretation of the results of the calculations and their consequences. Section~\ref{instblty} describes our suggestions on a possible phase diagram. We conclude in Section~\ref{sec:discussion} with a general discussion on how our results fit in with various other related problems and theoretical descriptions of non-fermi liquid metals. Various appendices contain details of calculations.

\section{Preliminaries}
\label{sec:prelim}
As mentioned above the low energy physics  is correctly described by focusing attention on Fermi surface patches with parallel normals\cite{polchinski,aliomil,dbl,maxsubir}.
This is because the interactions mediated by the boson field
are predominantly small-angle scattering processes. Furthermore short range four fermion interactions that couple different patches become unimportant at low energies\cite{polchinski,aliomil,dbl} as can be checked a posteriori after the two patch theory is solved. Thus the universal low energy physics of the  system is correctly captured by a theory that focuses on two opposite patches of the Fermi surface.
We focus henceforth on two opposite patches of Fermi surface; there are a number of subtle and important points related to the patch construction
that we elaborate on in Appendix \ref{app:anomalies}. The patch construction also has a number of immediate consequences for the behavior of many physical properties. These will be discussed in Section \ref{sec:physics}.

\begin{figure}[h]
\includegraphics[width=25mm]{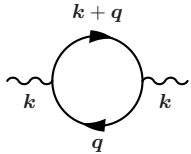}
\caption{1-loop boson self-energy.}
\label{fig.1loopboson}
\end{figure}

\begin{figure}[h]
\includegraphics[width=25mm]{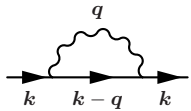}
\caption{1-loop fermion self-energy.}
\label{fig.1loopfermion}
\end{figure}

Consider patches of the Fermi surface with normals along ${\pm x}$.
We will denote the corresponding fermion fields $f_{R/L}$ where $R$ denotes the right patch and $L$ the left one.
It is useful to begin by considering the boson and fermion propagators in perturbation theory keeping just the leading one loop diagrams (Figs.~\ref{fig.1loopboson} and \ref{fig.1loopfermion}). The imaginary frequency boson propagator $D(\vec k, \omega)$ becomes
\begin{equation}
\label{bosonp}
D(\vec k, \omega)= \frac{1}{\gamma \frac{|\omega|}{|k_y|} + \frac{|k|^{z_b - 1}}{e^2}}
\end{equation}
with\footnote
{
For the case of the nematic,
we absorb the
dependence on the angle between the nematic ordering vector
and the patch in question into the coupling $e$.
This coupling $e$ specifies an energy scale below which
our low energy theory is applicable.
Since this energy scale vanishes at the `cold spots',
we must restrict attention to a patch of the Fermi surface
away from this direction.
}
$\gamma= \frac{1}{4\pi}$. Unless otherwise mentioned we will henceforth set $e = 1$.
The fermion propagator is determined by its self energy which at one-loop level takes the form
\begin{equation}
\label{sigma1loop}
\Sigma = -i \frac{1}{\lambda N} \sgn(\omega) |\omega|^{\frac{2}{z_b}}
\end{equation}
The constant $\lambda$ is given by
\begin{equation}
\lambda = 4\pi \sin\frac{2\pi}{z_b} \gamma^{\frac{z_b - 2}{z_b}}
\end{equation}
and thus vanishes linearly as $z_b \rightarrow 2$.
In terms of the scaling form in Eqn.~\ref{frmscl} this implies the fermionic dynamical critical exponent $z = \frac{z_b}{2}$ and $\alpha = 1$ as promised.

\vspace{1cm}

The arguments of Ref.~\onlinecite{ssl} show that a minimal Euclidean action that enables correct description of the low energy physics is given by
\begin{eqnarray}
S & = & S_f + S_{int} + S_a \\
S_f & = & \int d^2x d\tau \sum_{s\alpha} \bar{f}_{s\alpha} \left(\eta \partial_\tau -is \partial_x - \partial_y^2 \right) f_{s\alpha} \\
S_{int} & = & \int d^2x d\tau \frac{s}{\sqrt{N}}a \bar{f}_{s\alpha} f_{s\alpha} \\
S_a & = & \int_{\vec k, \omega} |k_y|^{z_b - 1} |\vec a(\vec k, \omega)|^2
\end{eqnarray}
Here $s = +1$ for the patch $R$ and $-1$ for the patch $L$. The parameter $\eta$ is taken to be small and positive.
 The field $a$ represents just the $x$-component of the vector field $a_i$. Indeed it is just this component that couples strongly to the patches with normals along $\pm x$. Note in particular that the boson field couples with {\em opposite} sign to the two antipodal patches. If on the other hand we were interested in the critical theory for a Pomeranchuk transition (such as a transition to an electronic $d$-wave nematic state in a two dimensional metal which microscopically has square lattice symmetry), the minimal action will have a very similar form except that the boson will couple with the {\em same} sign to antipodal patches. While this difference is unimportant for some properties it plays a crucial role in others. For instance the structure of the $2K_f$ singularities is completely altered between the gauge field and nematic models.

\section{One patch theory}
\label{sec:scaling}
We begin by focusing attention only on one patch, say the right one, and completely ignoring the other one. Indeed Ref.~\onlinecite{ssl} showed that the standard large-$N$ expansion leads to an apparently strongly coupled theory already in this simplified model. The main point is that a high loop diagram may formally look like it is high order in the $1/N$ expansion. However for many such diagrams the corresponding loop integral  diverges in the $\eta \rightarrow 0$ limit. This divergence may be regularized by using the one loop self energy in the fermion propagator. As this is of order $1/N$, the singular $\eta$ dependence is traded for an enhanced power of $N$ in the numerator. Consequently the naive $1/N$ counting is modified and an infinite number of diagrams survive in each order of $1/N$. A systematic way to keep track of the true power of $1/N$ is obtained by using a ``double-line" representation for the boson field that was previously used in the treatment of the electron-phonon interaction in metals\cite{shankarRG,phonondoubleline}.
It was shown that the $1/N$ expansion could be organized as a genus expansion with all ``planar" diagrams surviving to leading order.
Ref.~\onlinecite{ssl} further established that
 in the large-$N$ limit
the boson propagator is unrenormalized beyond 1-loop - in other words all higher loop diagrams that survive in the large-$N$ limit give vanishing contributions.
Each individual term contributing to the fermion self energy is (if one calculates using the 1-loop fermion propagator) finite, and has the same functional form as the 1-loop self energy: formally (at $z_b = 3$),
\begin{equation}
\label{SiglrgN}
\Sigma = - i \frac{1}{\lambda}\sgn(\omega)|\omega|^{\frac{2}{3}}\sum_n \frac{a_n N^{n-1}}{N^n}
\end{equation}
The $nth$ term in the sum comes from  diagrams that are formally of order $1/N^n$ in the large-$N$ expansion. However for all planar diagrams there is a compensating enhancement factor $N^{n-1}$ in the numerator so that each term is of order $1/N$.
The worry is whether the sum over the infinite contributing diagrams leads to something singular or not.

It is straightforward to see that these results carry over to general $z_b$. Indeed the kinematics leading to the divergences in the small $\eta$ limit depend only on the existence of the gapless Fermi surface and not on the detailed form of the boson propagator. When the divergence is regularized with the one loop fermion self energy, every $\frac{1}{\eta}$ is traded for a factor $\lambda N$. Eqn.~\ref{SiglrgN} is accordingly modified to
\begin{equation}
\Sigma = - i \frac{1}{\lambda} \sgn(\omega)|\omega|^{\frac{2}{z_b}}\sum_n \frac{b_n \left(\lambda N\right)^{n-1}}{N^n}
\end{equation}
Here the coefficients $b_n$ are all independent of $N$ but in general depend on $z_b$.
The utility of the small $z_b - 2$ limit where $\lambda \propto z_b - 2$ is now apparent. So long as $z_b - 2$ is of order $1/N$, the enhancement factor $\left(\lambda N \right)^{n-1}$ in the numerator of each term above is finite. If further the $b_n$'s are sufficiently non-singular when $z_b \rightarrow 2$ then in the large-$N$ limit only the $n = 1$ term survives. This is just the one loop answer which is thus exact in this limit.
This claim can be illustrated explicitly by calculating a particular instance of a dangerous diagram which contributes to the series above at $n = 2$, such as the one shown in Fig.~\ref{fig.3loop}.  We do this in Appendix \ref{app:1patchdia} and show that though it is of order $1/N$ for general $z_b$, in the limit $z_b - 2 \propto 1/N$ it becomes higher order in $1/N$.  In particular the corresponding coefficient $b_2$ has a finite limit as $z_b \rightarrow 2$ so that its contribution is of order $1/N^2$.
This is in fact expected to be true for all the $b_n$ which have limits when $z_b \rightarrow 2$ such that to leading order high $n$ terms in the series give subdominant powers of $N$ to the leading order result. Indeed the absence of $\omega \log^2\omega$ terms\cite{hlr,aliomil,adybert} in the self energy exactly at $z_b = 2$ implies that the self energy for small $\lambda$ at fixed $N$ has at most one inverse power of $\lambda$. Furthermore in this limit the self energy can be explicitly calculated using a perturbative RG technique (see Section~\ref{pertRG}). The answer agrees exactly with the leading order term in the series above. This means that the small $z_b -2$ behavior of $b_n$ is such as to keep the high $n$ terms subdominant to the leading order one in the large-$N$ limit.

The boson propagator is also given exactly by the one-loop answer. For arbitrary $z_b$ this follows from the arguments of Ref.~\onlinecite{ssl} due to the vanishing of higher order planar diagrams. However in the small $z_b - 2$ limit it also follows for the same reason as above - the enhancement factors that render higher order diagrams to be of the same nominal order in $1/N$ all become innocuous in the small $z_b - 2$ limit.

We emphasize that even though we have used the formal device of small $z_b - 2$ to control the large-$N$ expansion the frequency dependence of the
fermion self energy $\Sigma \propto -i \sgn(\omega)|\omega|^{\frac{2}{z_b}}$ is exact to all orders in $z_b - 2$. The smallness of $z_b -2$ merely assures us that the proportionality constant has a sensible $1/N$ expansion.

Similar results also apply to the fermion-boson vertex which  to leading order in $1/N$ is unrenormalized.

It is useful to understand the scaling structure of the low energy physics described above. We have shown that in the one patch theory the large-$N$, small $z_b - 2$ limit, the low energy
physics is described by a fixed point invariant under the scaling transformation

\begin{eqnarray}
\label{scaling}
\omega' & = & \omega b^{\frac{z_b}{2}} \\
p_x' & = & p_x b \\
p_y' & = & p_y' b^{\frac{1}{2}} \\
f'_{\alpha s}(p_x', p_y', \omega') & = & b^{-\frac{z_b + 5}{4}} f_{\alpha s}(p_x, p_y, \omega) \\
\label{lastscaling}
a'(p_x', p_y',  \omega') & = & b^{-\frac{1+z_b}{2}}a(p_x, p_y, \omega).
\end{eqnarray}
This is the same scaling structure that is obtained in a naive one loop approximation. Though Ref.~\onlinecite{ssl} has raised concerns over whether this scaling is internally consistent in the large-$N$ limit we see from the preceding analysis that it indeed is if the limit of small $z_b - 2$ is also simultaneously taken.

The considerations of Ref.~\onlinecite{aliomil} can now be used to argue that this scaling structure is exact to all orders in the $1/N$ expansion (so long as $z_b - 2$ is small) within this one patch theory. More specifically the boson propagator and fermion self energy will have the same functional forms (Eqn.~\ref{bosonp} and Eqn.~\ref{sigma1loop}) as the one loop answers. Thus we expect that in the $N$, $z_b$ plane there is a region of finite extent where the scaling structure above is preserved. This conclusion is further bolstered by the perturbative RG analysis for finite $N$, small $z_b -2$ (see Section \ref{pertRG}).

\section{Two patches}
\label{sec:twopatch}
The considerations above are readily generalized to the two patch theory. Indeed as before at each order of $1/N$ there are a number of diagrams that
are divergent in the small $\eta$ limit. This is traded for an enhancement factor of $\lambda N$ raised to some power when the one-loop self energy is used instead. For $z_b - 2$ of order $1/N$ these enhancement factors become finite and a controlled $1/N$ expansion emerges. The structure of the leading order contribution in $1/N$ to the fermion self energy, the boson propagator, or the interaction vertex is then not modified from the one patch theory, and thus retains its RPA form. In particular the scaling structure in Eqns.~\ref{scaling}-\ref{lastscaling} is preserved. Below we examine higher order corrections in the $1/N$ expansion, and show that new singularities appear at $o(1/N^2)$.  These could potentially modify the scaling structure from that in Eqns.~\ref{scaling}-\ref{lastscaling} above. However we show that in the gauge field model these contribute only to subdominant corrections to the one loop fermion Green's function. On the other hand for the nematic transition there is indeed a modification of the fermion scaling `dimension' at $o(1/N^2)$. Our calculations rely on the very recent impressive results of Metlitski and Sachdev\cite{maxsubir} who studied the boson propagator and fermion self energy in a direct (albeit uncontrolled) perturbative loop expansion upto three loop order. Here we will show that calculations along the lines of those in
Ref.~\onlinecite{maxsubir} leads to controlled results for various physical quantities within our modified $1/N$ expansion.

\begin{figure}[ht]
\centering
\includegraphics[width=40mm]{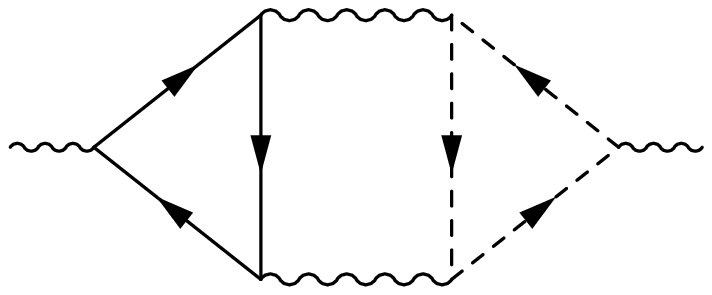}
\hspace{4mm}
\includegraphics[width=40mm]{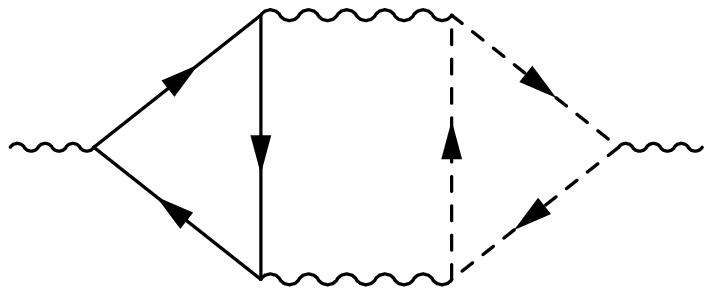}
\caption{Three-loop boson self-energy diagrams.}
\label{fig.aslamazov}
\end{figure}

First consider the boson propagator. At $z_b = 3$ the analysis of Ref.~\onlinecite{maxsubir} established that to three loop order there is no shift of the true dynamical critical exponent. This was done by showing that the boson propagator at zero external frequency stays proportional to $q^2$ up to three loops. However the three loop contribution is of order $\sqrt{N}$ bigger than the one loop contribution thereby casting doubts on the existence of a sensible large-$N$ limit. We will suggest an interpretation of this result in Section \ref{instblty}. But for now we discuss the results of an identical analysis in our limit of small $z_b-2 \sim o(1/N)$. First we note that the leading order term in the inverse gauge propagator is of order $1$ with our conventions. Next higher loop diagrams clearly give subdominant powers of $1/N$ as there are no enhancement factors in the limit of small $z_b - 2$. At zero external frequency the leading $1/N$ correction comes from the two diagrams shown in Fig.~\ref{fig.aslamazov}. These same diagrams were calculated in Ref.~\onlinecite{maxsubir} at $z_b = 3$. Repeating for general $z_b$, we find the $1/N$ correction
\begin{equation}
\pm c\frac{f_1(\lambda N)}{N} |q_y|^{z_b -1}
\end{equation}
with the $-$ sign for the nematic critical point and the $+$ sign for the gauge model. The function $f_1$ is evaluated in Appendix \ref{app:3loop} , and is readily seen to have a finite limit when $z_b \rightarrow 2$. For large $\lambda N$, we have $f_1(\lambda N) \sim \left(\lambda N \right)^{\frac{z_b}{2}}$ in agreement with the result of Ref.~\onlinecite{maxsubir} when $z_b = 3$. We see explicitly that when $z_b - 2$ is $o(1/N)$ these three loop contributions are down by a factor $1/N$ compared to the one loop term. Thus the large-$N$ expansion is indeed well defined in this limit.

\begin{figure}[h]
\centering
\includegraphics[width=30mm]{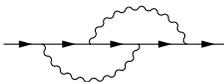}
\caption{Two-loop fermion self-energy diagrams
merely renormalize the coefficient of $|\omega|^{2/z_b}$.
}
\label{fig.2loopfermion}
\end{figure}

\begin{figure}[h]
\centering
\includegraphics[width=42mm]{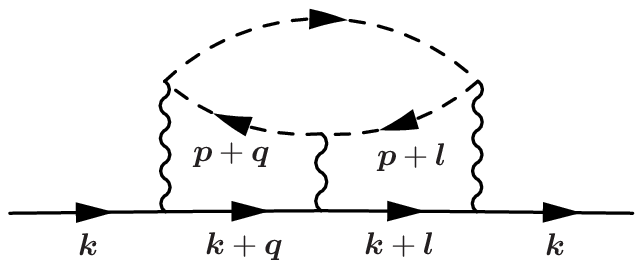}
\includegraphics[width=42mm]{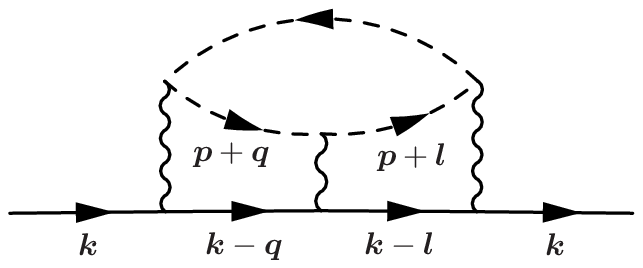}
\caption{Three-loop Fermion self-energy diagrams involving fermions on both patches. Fermions on the right patch are denoted by solid lines, fermions on the left patch by dashed lines.}
\label{fig.3looplr}
\end{figure}

Next we consider the fermion propagator. The one-loop self energy is inversely proportional to $\lambda N$ and hence is of order $1$. It is easy to see by explicit calculation that the two loop diagram shown in Fig.~\ref{fig.2loopfermion} is momentum independent, and merely provides an $o(1/N^2)$ modification of the coefficient of the frequency dependent part of the self energy. The most important effect at this order comes from the two graphs shown in Fig.~\ref{fig.3looplr}. At zero external frequency these graphs lead to singular momentum dependence\cite{maxsubir}. Details are given in  Appendix \ref{app:3loop}. We find (for the right moving fermion)
\begin{equation}
\delta \Sigma (\vec p, \omega = 0) = \pm \frac{4}{3N^2} J(\lambda N) (p_x + p_y^2)\ln\left(\frac{\Lambda}{\left(p_x + p_y^2 \right)^{\frac{z_b}{2}}}\right)\label{eqn:jequation}
\end{equation}
where the function $J(\lambda N)$ is defined in Appendix \ref{app:3loop}
and is positive definite.
The $+$ sign applies to the nematic critical point and the $-$ sign to the gauge model.
In a renormalization group framework this can be interpreted as the leading terms of a singular contribution to the self energy of the form
\begin{equation}
\label{singself}
(p_x + p_y^2)^{1 \mp \frac{4}{3N^2} J(\lambda N)}
\end{equation}
A similar contribution exists in the frequency dependent part as well, consistent with the dynamical scaling. In the gauge field case the plus sign applies, and this singular correction is subdominant to the terms that already exist at leading order. Indeed we expect that any perturbation of the fixed point by irrelevant perturbations will generate an analytic contribution to the momentum dependence of the self energy that will then dominate over the singular corrections found at $o(1/N^2)$. Though the leading order frequency dependence is not analytic we expect that if we use the large-$N$ fermion propagators to calculate the effects of an irrelevant operator in perturbation theory we will simply again generate a $|\omega|^{\frac{2}{z_b}}$ term. This will dominate over the singular order $1/N^2$ correction. Thus we conclude that in the gauge model the leading singularities are correctly given by the RPA forms at least to order $1/N^2$. We note that our interpretation is different from that in Ref.~\onlinecite{maxsubir}.

In the nematic case the minus sign applies in the exponent of Eqn.~\ref{singself}. This is more singular than the `bare' momentum dependence of the inverse Green's function, and consequently will dominate the low energy physics near the Fermi surface. This singular correction can be interpreted as a shift of the scaling of the fermion fields from that in
Eqns.~\ref{scaling}-\ref{lastscaling}. Thus to order $1/N^2$ we have
\begin{equation}
f'_{\alpha s}(p_x', p_y', \omega')  =  b^{-\frac{z_b + 5 - \eta_f}{4}} f_{\alpha s}(p_x, p_y, \omega)
\end{equation}
where
\begin{equation}
\label{etaf}
\eta_f = \frac{4}{3N^2} J(\lambda N)
\end{equation}
 All the other scaling equations remain unmodified. This implies that the fermion Green's function satisfies the scaling form with $\alpha = 1- \eta_f$.

What is the physical origin of the signs and the differences between the gauge and nematic models? We explain this in Section \ref{sec:physics}. To set the stage we first calculate singularities in some other quantities within the general two patch theory.

\section{$2K_f$ and other singularities}
\label{twokf}
 The calculations of Ref.~\onlinecite{aliomil} on the gauge model showed that the response to an external field that couples to the fermion density at momentum $2K_f$ is modified from that of a Fermi liquid due to the gauge interaction. The physical origin of this effect is clear. First there is a suppression of the $2K_f$ response coming from the smearing of the Landau quasiparticle due to the boson interaction. Second there is (in the gauge model) an enhancement coming from the ``Amperean" attraction between a particle at $K_f$ and a hole at $-K_f$. The gauge currents of such a particle and hole are parallel to each other so that the gauge interaction is attractive for such a particle-hole pair. The net modification of the $2K_f$ singularity is determined by the interplay between these two effects. We emphasize that the Amperean enhancement is specific to the gauge field problem. For the closely analogous problem of a quantum critical point associated with a Pomeranchuk transition, the particle-hole interaction mediated by order parameter fluctuations is repulsive. This goes in the same direction as the effect due to the smearing of the Landau quasiparticle. So there is no competition and we expect that the $2K_f$ singularities are simply suppressed when compared with the Fermi liquid at the Pomeranchuk transition.

The structure of the fixed point in the large-$N$, small $z_b - 2$ limit enables a controlled calculation of these effects. Consider an external field that couples to the $2K_f$ fermion density through the following term in the action
\begin{equation}
\label{S_u}
u\int d^2x d\tau \bar{f}_L f_R + h.c
\end{equation}
By power counting it is easy to see that the scaling in Eqns.~\ref{scaling}-\ref{lastscaling} implies that the coupling $u$ scales as
\begin{equation}
u' = u b
\end{equation}
In general this will be modified at order $1/N$ as we demonstrate below. But first let us understand how the scaling of the $u$ determines the $2K_f$ singularities. Assume in general that $u$ scales as
\begin{equation}
u' = u b^{\phi_u}
\end{equation}
This implies that the operator $\rho_{2K_f}(\vec x, \tau) = \bar{f}_L (\vec x, \tau) f_R (\vec x, \tau)$ scales as
\begin{equation}
\rho'_{2K_f}(x', y', \tau') = b^{\Delta} \rho_{2k_f}(x,y,\tau),
\end{equation}
with
\begin{equation}
\label{Delta}
\Delta = \frac{z_b + 3}{2} - \phi_u .
\end{equation}
This determines the behavior of the singular part of the $2K_f$ density correlation function $C_{2K_f}(x, y, \tau) = \langle \rho_{2K_f}^*(x,y,\tau) \rho_{2K_f}(0,0,0) \rangle$. Its Fourier transform satisfies
\begin{equation}
C_{2K_f}(p_x, p_y, \omega) = b^{\frac{3 + z_b}{2} - 2\Delta} C'_{2K_f} \left(p_x', p_y', \omega' \right)
\end{equation}
Note that the momenta $p_x, p_y$ describe the {\em deviation} of the full momentum from $2K_f \hat{x}$ in this correlation function. We then immediately have the scaling form
\begin{equation}
\label{2kfscaling}
C_{2K_f} (p_x, p_y, \omega) = \frac{1}{\omega^{1+ \frac{3 - 4\Delta}{z_b}}} {\cal C}\left(\frac{\omega}{|p_y|^{z_b}}, \frac{p_x}{p_y^2}\right)
\end{equation}
Note that in the usual Fermi liquid case, we have $z_b = 2$, $\phi_u = 1$ which reproduces the well known square root frequency dependence of the singular part of the $2K_f$ correlations.

The leading order correction to $\phi_u$ in the $1/N$ expansion comes from the one loop vertex correction diagram of Fig.~\ref{fig.vertex}. To calculate it we combine the large-$N$ expansion with an RG transformation in which internal loop integrals are performed over a thin shell in $(\vec p, \omega)$ space. It will be convenient to define the RG so that we
integrate over arbitrary $p_x, \omega$ but over a shell in $p_y$ where $\Lambda > |p_y| > \Lambda/ b^{\frac{1}{2}}$. The vertex correction becomes
\begin{equation}
\delta u = -\frac{1}{N} \int_{\vec p, \omega} D(p_y, \omega) {\cal G}_R(p_x, p_y, \omega) {\cal G}_L(p_x, p_y, \omega)
\end{equation}
Here ${\cal G}_{R/L}$ are the propagators of the right and left moving fermions respectively. These propagators include the singular frequency dependent self energy discussed in previous sections. Note the minus sign in front which comes from the different signs with which the gauge field couples to the left and right fermions. In contrast at a Pomeranchuk transition the minus sign will be absent. Consequently what is an enhanced vertex in the gauge field problem will become a suppressed vertex at the Pomeranchuk transition.

We evaluate the integral in Appendix \ref{app:vertex}. The result takes the form
\begin{equation}
\label{vertex}
\delta u = \frac{1}{4\pi^2 N} g(\lambda N, z_b) \ln(b)
\end{equation}
where the function $g$ is given by
\begin{equation}
g(v, z_b) = \int_0^\infty dt \left(\frac{1}{\gamma t + 1} \right)\frac{v t^{\frac{2}{z_b}}}{t^{\frac{4}{z_b}} + v^2}
\end{equation}
where $\gamma = \frac{1}{4\pi}$.
For fixed $v$ in the limit that $z_b =2$ this function has a sensible limit which we demote $g_2(v)$. We find
\begin{equation}
g_2(v) = \frac{\pi \gamma v^2}{2(\gamma^2 v^2 + 1)}\left( 1 - \frac{2}{\pi \gamma v} \ln(\gamma v)\right)
\end{equation}

The vertex correction in Eqn.~\ref{vertex} implies a modified scaling exponent for $u$:
\begin{equation}
\phi_u = 1 + \frac{1}{4\pi^2 N} g(\lambda N, z_b)
\end{equation}
As $g$ has a finite limit when $z_b \rightarrow 2$, $N \rightarrow \infty$ but $\lambda N$ is finite, this modification is of order $1/N$. We note that it is sufficient to evaluate $g$ at $z_b = 2$ in this limit. To leading order in $\epsilon = z_b - 2, \frac{1}{N}$, we therefore have (using  $\lambda = 2\pi^2 \epsilon$ appropriate for small $\epsilon$)
\begin{equation}
\phi_u = 1 + \frac{1}{4\pi^2 N} g_2(2\pi^2 \epsilon N).
\end{equation}
Inserting this into Eqns.~\ref{Delta} and \ref{2kfscaling} we see that for small $\epsilon, 1/N$
the power of $\omega$ in the $2K_f$ singularity
is
\be
\label{eq:ampere}
\frac{1}{2} + \frac{\epsilon}{4}\( 1 - \frac{g_2(2\pi^2 \epsilon N)}{\pi \epsilon N} \) .\ee
Since $g_2(v)/v$ runs from $\infty$ to $0$ as $v$ runs from $0$ to $\infty$,
the $2K_f$ singularity
is suppressed over that of the Fermi liquid for $\epsilon N \rightarrow \infty$ while it is enhanced when $\epsilon N \rightarrow 0$.
The competition
in Eqn.~\ref{eq:ampere}
between the first and second term in parentheses is
precisely the competition between quasiparticle smearing and
Amperean attraction described at the beginning of this section.

\bigskip
The structure of the singularities in the Cooper channel will also be modified from that of the Fermi liquid. But here the Amperean interaction is repulsive in the gauge field case and thus the Cooper singularities will be weaker than in the Fermi liquid. On the other hand at the nematic quantum critical point the boson mediated interaction is attractive in the Cooper channel. Consequently there will be an enhancement of the pairing vertex due to boson exchange. This leads to an enhancement of the Cooper singularities compared with the Fermi liquid.   A complete discussion of this effect is more complicated than the $2K_f$ singularities and will be presented elsewhere\cite{dmrossunpub}.
Within the large-$N$ expansion for small $z_b - 2$, in contrast to the $2K_f$ singularity, there is an $o(1)$ correction to the tree level scaling exponent of the Cooper vertex which receives contributions from many diagrams\cite{dmrossunpub}. Quantitative calculation of this correction as well as a study of the interesting interplay between superconductivity and critical nematic fluctuations will appear elsewhere\cite{dmrossunpub}.

\section{Perturbative fixed point for finite $N$, small $\epsilon$ and $2K_f$ singularities}
\label{pertRG}
A different controlled limit was previously discussed in Ref.~\onlinecite{chetan}. This is obtained by considering finite $N$, and small $\epsilon$ where there is a perturbatively accessible renormalization group fermi-liquid fixed point.
In this section we study the $2K_f$ singularities at this fixed point.   It will be useful for our purposes to define a RG scheme that is slightly different from Ref.~\onlinecite{chetan}. So let us first reproduce their main result. Consider the two patch action
\begin{eqnarray}
S & = & S_f + S_{int} + S_a \\
S_f & = & \int d^2x d\tau \sum_{s\alpha} \bar{f}_{s\alpha} \left(\partial_\tau -is \partial_x - \partial_y^2 \right) f_{s\alpha} \\
S_{int} & = & \int d^2x d\tau \frac{s e}{\sqrt{N}}a \bar{f}_{s\alpha} f_{s\alpha} \\
S_a & = & \int_{\vec k, \omega} |k_y|^{z_b - 1} |\vec a(\vec k, \omega)|^2 + .....
\end{eqnarray}
Compared with the `minimal' action of Section \ref{sec:prelim} above, we have set $\eta = 1$ and have reinstated the boson coupling $e$. The ellipses in the last equation refer to other operators irrelevant at the RG fixed point to be described below (for instance an $\omega^2$ term in the quadratic gauge action). When $e = 0$, the fermions and boson field are described by two decoupled Gaussian theories. This is at a fixed point under the scaling
\begin{eqnarray}
\tau' & = & \frac{\tau}{b} \\
x' & = & \frac{x}{b} \\
y' & = & \frac{y}{\sqrt{b}} \\
f'(x',y', \tau') & = & b^{\frac{3}{4}}f(x,y,\tau) \\
a'(x',y', \tau') & = & b^{\frac{3}{2} - \frac{z_b}{4}}a(x,y,\tau)
\end{eqnarray}
Now turn on a small $e \neq 0$. By power counting we find
\begin{equation}
e' = e b^{\frac{z_b - 2}{4}}
\end{equation}
In differential form (if we let $ b = 1 + dl$) we get
\begin{equation}
\frac{de^2}{dl} = \frac{\epsilon}{2} e^2
\end{equation}
Thus $e^2$ is relevant for $\epsilon > 0$ and irrelevant for $\epsilon < 0$. To determine the fate of the theory for $\epsilon > 0$ let us study the one loop beta function for $e^2$. We will define the RG by integrating out modes with $\Lambda > |q_y| > \frac{\Lambda}{\sqrt{b}}$. To order $e^2$, the fermion self energy is given by the integral
\begin{equation}
\Sigma  =  \frac{e^2}{4\pi^3 N} \int_{q_x, \omega'}\int^{\Lambda}_{\frac{\Lambda}{\sqrt{b}}}dq_y D(q_y, \omega'){\cal G}(\vec p-\vec q, \omega-\omega')
\end{equation}
where $D$ is the gauge propagator in the bare action.  After doing the $q_x$ integral the remaining $\omega'$ integral only gets contributions from small $\omega'$ so we can replace the gauge propagator by its low frequency form $\frac{1}{|q_y|^{z_b - 1}}$. For small external frequancy $\omega$ we get
\begin{equation}
\Sigma =  -\frac{ie^2 \omega}{2\pi^2 N}\int^{\Lambda}_{\frac{\Lambda}{\sqrt{b}}}\frac{dq_y}{q_y^{z_b -1}}
\end{equation}
Anticipating that there is a new fixed point when $e^2/N \sim o(\epsilon)$, we replace the integrand by its value at $z_b = 2$ to get
\begin{equation}
\Sigma = -\frac{ie^2 \omega}{4\pi^2 N} \ln b
\end{equation}
Thus the inverse fermion propagator takes the form
\begin{equation}
i\omega \left( 1 + \frac{e^2}{4\pi^2 N} \ln b \right) - p_x - p_y^2
\end{equation}
\begin{figure}[h]
\centering
\includegraphics[width=25mm]{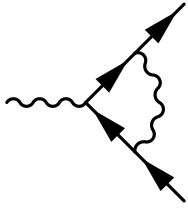}
\caption{One loop correction to the fermion-boson vertex.}
\label{fig.bfvertex}
\end{figure}
The correction to the fermion-boson vertex (see Fig.~\ref{fig.bfvertex}) at order $e^2$ vanishes as the $q_x$ integral has poles only on one side of the complex plane. Finally the change to the boson propagator is also zero if only modes with high $|q_y|$ are integrated out. Thus the only change is in $\omega$ dependence of the fermion propagator. This may be incorporated into a modified scaling
\begin{eqnarray}
\omega' & = & \omega \left(  1 + \frac{e^2 }{4\pi^2 N} \ln b \right) \\
& = & \omega b^{1 + \frac{e^2 }{4\pi^2 N}}
\end{eqnarray}
This implies
\begin{eqnarray}
\tau' & = & \frac{\tau}{b^{1 + \frac{e^2}{4\pi^2 N}}} \\
 x' & = & \frac{x}{b} \\
y' & = & \frac{y}{\sqrt{b}} \\
f'(x',y', \tau') & = & b^{\frac{3}{4} + \frac{e^2}{8\pi^2 N}}f(x,y,\tau) \\
a'(x',y', \tau') & = & b^{\frac{3}{2} - \frac{z_b}{4} + \frac{e^2}{8\pi^2 N}}a(x,y,\tau)
\end{eqnarray}
The modification to the flow of the coupling $e$ is now readily obtained to be
\begin{equation}
e' = e b^{\frac{z_b - 2}{4} - \frac{e^2}{8\pi^2 N}}
\end{equation}
In differential form this implies the flow equation
\begin{equation}
\frac{de^2}{dl} = \frac{\epsilon e^2}{2} - \frac{e^4}{4\pi^2 N}
\end{equation}
Thus we indeed find a fixed point when
\begin{equation}
e^2_* = 2\pi^2 N \epsilon
\end{equation}
Right at the fixed point the scaling equations above are identical (to within order $\epsilon$) to those found earlier in Section \ref{sec:scaling} and indeed to that expected based on RPA. The differences from RPA discussed in earlier sections in the fermion propagator come from three loop calculations, and hence are not expected to show up till order $\epsilon^3$.

The singularities of many physical quantities can be usefully calculated within this $\epsilon$ expansion and provides an alternate controlled limit to the one we have discussed. As an illustration let us calculate the boson propagator and the fermion self energy. Let the bare value of the electric charge at the cut-off scale be $e_0$. The boson propagator is given by the usual one loop diagram and takes the form
\begin{equation}
\label{bosonprg}
D(q_y, \omega) = \frac{1}{\frac{e_0^2}{4\pi}\frac{|\omega|}{|q_y|} + |q_y|^{1+ \epsilon}}
\end{equation}
The Landau damping term does not acquire any corrections from this perturbative answer at least upto the order to which the RG has been performed. To obtain the frequency dependence of the fermion propagator we examine the flow of the coefficient of the $i\omega$ term calculated above. Let us denote this coefficient $\eta(l)$ at an RG scale $l$. The calculation above gives the flow equation
\begin{equation}
\frac{d\eta}{dl} = \frac{\eta e^2}{4\pi^2N}
\end{equation}
Combining with the equation for $e^2$ we obtain
\begin{equation}
\frac{d\left(\eta e^2\right)}{dl} = \frac{\epsilon}{2} \eta e^2
\end{equation}
Thus we find
\begin{equation}
\eta(l) e^2(l) = e_0^2 e^{\frac{\epsilon l}{2}}
\end{equation}
where we set $\eta(l = 0) = 1$, and $e_0^2$ is the bare coupling at the cut-off scale. The frequency dependence of the fermion self energy is then obtained by setting $l = \ln\frac{\Lambda^2}{\omega}$. In the limit $\omega \rightarrow 0$, we may set $e^2(l) = e_*^2$ so that
\begin{equation}
\eta(\omega) = \frac{e_0^2}{2\pi^2 N \epsilon} \left(\frac{\Lambda^2}{\omega}\right)^{\frac{\epsilon}{2}}
\end{equation}
Then at order $\epsilon$ the self energy becomes
\begin{equation}
\label{sigmarg}
\Sigma(\omega) =-i \frac{e_0^2}{2\pi^2 N \epsilon} |\omega|^{1 - \frac{\epsilon}{2}}\sgn(\omega)
\end{equation}
Here we have ignored a term $\Lambda^\epsilon$ in the numerator to this order in $\epsilon$.

To compare with the results of previous sections we need to set the bare coupling $e_0^2   = 1$. Note in particular that the prefactor to the frequency dependence is exactly $\frac{1}{\lambda N}$ consistent with the earlier analysis. On the other hand to calculate the scaling dimensions of any operator directly within this epsilon expansion we need to sit right at the fixed point and perturb the theory with that operator. The fixed point theory corresponds to setting the bare coupling $e_0 = e_*$. Thus the boson propagator and fermion self energy right at the fixed point take the forms
\begin{eqnarray}
\label{dsfp}
D_*(q_y, \omega) & = & \frac{1}{\frac{e_*^2}{4\pi}\frac{|\omega|}{|q_y|} + |q_y|^{1+ \epsilon}} \\
\Sigma_*(\omega) & = & -i   |\omega|^{1 - \frac{\epsilon}{2}}\sgn(\omega)
\label{dsfplast}
\end{eqnarray}
Note that as $\epsilon \rightarrow 0$ these fixed point propagators go over smoothly into those of the `decoupled' Gaussian fixed point, as indeed they must.

As it does not seem to be available in the literature,  let us now calculate the scaling exponent for the $2K_f$ singularity at this $o(\epsilon)$ fixed point. As before we add the term in Eqn.~\ref{S_u} to the action. By power counting we again find
\begin{equation}
u' = ub
\end{equation}
This is modified at leading order of $\epsilon$ by the same vertex correction diagram as before. Evaluating the integral as before we find
\begin{equation}
\delta u = \frac{e_*^2 u}{4\pi^2N}\int^{\Lambda}_{\frac{\Lambda}{\sqrt{b}}}dq_y\int_0^\infty d\omega D_*(q_y, \omega)\frac{\tilde{\Sigma}_*(\omega)}{\left(\tilde{\Sigma}_* (\omega) \right)^2 + q_y^4 }.
\end{equation}
Here we have written $\Sigma_*(\omega) = -i\tilde{\Sigma}_*(\omega)$. Naively as the vertex correction is already order $e^2 \sim \epsilon$ we should replace the integrand by its value at $\epsilon = 0$, {\em i.e} by the fermion and boson propagators at the Gaussian fixed point. However at the Gaussian fixed point $\tilde{\Sigma} = \omega$ and the frequency integral is logarithmically divergent at large $\omega$. This signals that the leading order $\epsilon$ correction to $\phi_u$ is not analytic in $\epsilon$. To extract it we keep the correct boson propagator and fermion self energy at the $o(\epsilon)$ fixed point calculated in Eqns.~\ref{dsfp}-~\ref{dsfplast}.

Inserting these into the integral for the vertex correction, we see that the high-$\omega$ divergence of the $\omega$ integral is cutoff by the presence of the Landau damping term in the boson propagator. The integral is readily evaluated for small $\epsilon$, and we find
\begin{equation}
\delta u = \frac{u \epsilon}{2} \ln\left(\frac{2}{\pi \epsilon N}\right) \ln b.
\end{equation}
Thus we get the modified scaling equation
\begin{equation}
u' = u b^{1 + \frac{\epsilon}{2} \ln\left(\frac{2}{\pi \epsilon N}\right)}.
\end{equation}
so that
\begin{equation}
\phi_u = 1 + \frac{\epsilon}{2} \ln\left(\frac{2}{\pi \epsilon N}\right)
\end{equation}
Inserting into Eqns.~\ref{Delta} and \ref{2kfscaling} we see that the $2K_f$ singularity is enhanced compared to the Fermi liquid at this fixed point. Note that this answer for $\phi_u$ agrees exactly with the result of Section \ref{twokf} when the limit of $\epsilon N \rightarrow 0$ is taken.

\section{Physical picture}
\label{sec:physics}
In this section we discuss some qualitative aspects of the physical picture of the low energy physics, and show how we may understand the results of some of the detailed calculations presented in previous sections. First we notice that at low energies the Fermi surface is sharp even though the Landau quasiparticle has been destroyed. This is qualitatively the same as in the RPA but at least in the nematic case the detailed singular structure is modified. Despite this, as argued in many previous papers\cite{hlr,ybkim,aliomil} this is a compressible state. This follows immediately from the general argument in Appendix \ref{app:anomalies} that the compressibility does not receive any singular contributions from the low energy scale invariant fluctuations. Hence this non-fermi liquid state has a  finite and non-zero compressibility. Actually in the nematic case changing the chemical potential will also in general drive the system away from the critical point. This leads to a singular contribution to the ground state energy as a function of chemical potential which could lead to a singular contribution to the compressibility\cite{maxsubir}. The safe statement then is that the differential change in density in response to a change in location within the phase boundary is finite.

Actually the patch construction implies an even stronger result. Consider the susceptibility to a deformation of the Fermi surface in any angular momentum channel ({\em i.e} the response to an external  field that couples to the corresponding shape distortion of the Fermi surface). The universal singularities in this quantity are obtained by examining this coupling within the patch construction. But the patch theory does not know anything about the angular dependence of the probe field. So within each patch this external field couples in the same way as an external chemical potential. Consequently (just as for the contribution to the compressibility from the low energy density fluctuations) there is no singular contribution to the susceptibility in any angular momentum channel which all stay finite and non-zero. For the nematic critical point the only exception is the order parameter channel itself ($l =2$ for the $d$-wave nematic). In that case this argument implies that the critical behavior of the order parameter susceptibility is correctly given by the mean field Hertz answer and receives no singular corrections from the fermions. These arguments provide a simple explanation of some recent results  for the nematic critical point obtained through detailed calculations\cite{chubukovfl}.  Consider the approach to the quantum critical point from the symmetric side where there is no nematic order. At low energies the corresponding metal is described by Fermi liquid theory characterized by Landau quasiparticles with an effective mass $m^*$ and various Landau parameters. On approaching the quantum critical point standard scaling arguments show that the effective mass diverges with an exponent as
\begin{equation}
m^* \sim \delta^{-\frac{z_b - 2}{z_b - 1}}
\end{equation}
In the Fermi liquid phase the compressibility is expressed  in terms of $m^*$ and the Landau parameter $F^0_s$ as
\begin{equation}
\kappa \propto \frac{m^*}{1+ F^0_s}
\end{equation}
The constancy of $\kappa$ as the critical point is approached implies that the Landau parameter $F^0_s$ diverges in exactly the same way as the effective mass. Applying this reasoning to other angular momentum channels we see that the Landau parameters in all angular momentum channels (except the order parameter one itself) must diverge in the same way as $m^*$ so as to give a constant susceptibility at the critical point. This is exactly the conclusion of Ref. \onlinecite{chubukovfl}.

In either the gauge model or the nematic critical point the patch construction implies that the only singular modification from RPA in the charge density response happens at the $2K_f$ wavevectors (modulo the caveat discussed above for the $q= 0$ response in the nematic case). As explained in detail in Section \ref{twokf} apart from the $2K_f$ particle-hole correlations the main modifications from the Fermi liquid in the two particle response are in the structure of the pairing correlations. Whether the $2K_f$ and pairing correlations are enhanced or not compared to the Fermi liquid is largely determined by the Amperean rules. In the gauge field case the pair correlations are suppressed and the $2K_f$ potentially enhanced while the opposite is true for the nematic critical point.

Consider now the $1/N^2$ calculation of the fermion self energy described in Section \ref{sec:twopatch}. The singular contribution to the self energy comes the two diagrams shown in Fig.~\ref{fig.3looplr}. We note that both diagrams may be expressed in terms of appropriate two-particle scattering amplitudes. Fig.~\ref{fig.cooper3} is a scattering amplitude in the (particle-particle) Cooper channel while Fig.~\ref{fig.2Kf3} is a scattering amplitude in the particle-hole $2K_f$ channel. Based on the physical picture dictated by the Amperean rules, we expect that in the nematic case the Cooper channel diagram by itself leads to a self energy that is more singular than the `bare' terms in the action while the $2K_f$ diagram by itself leads to a singularity that is less singular than the bare term.  The situation is clearly reversed in the gauge field model. This physical picture thus enables us to understand the signs of the contributions of the two diagrams in the calculation.

\begin{figure}
\centering
\includegraphics[width=60mm]{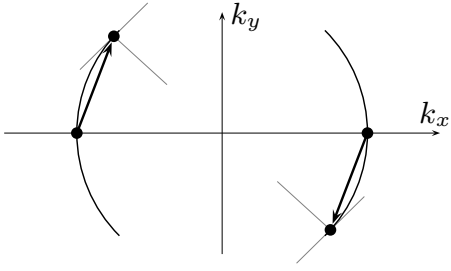}
\caption{After any scattering event in the Cooper channel, two fermions remain perfectly nested.}
\label{fig:nesting1}
\end{figure}

\begin{figure}
\centering
\includegraphics[width=60mm]{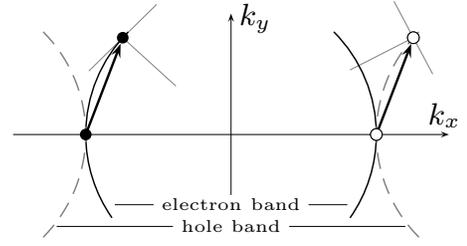}
\caption{After a scattering event in the 2$K_f$ channel, two fermions are no longer perfectly nested.}
\label{fig:nesting2}
\end{figure}



\begin{figure}
\centering
\includegraphics[width=60mm]{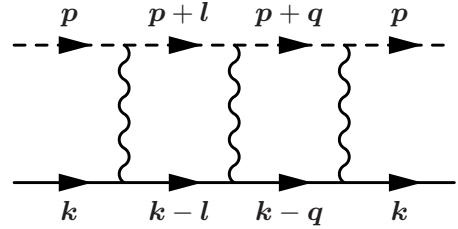}
\caption{The cooper-channel scattering amplitude.}
\label{fig.cooper3}
\end{figure}


\begin{figure}
\centering
\includegraphics[width=60mm]{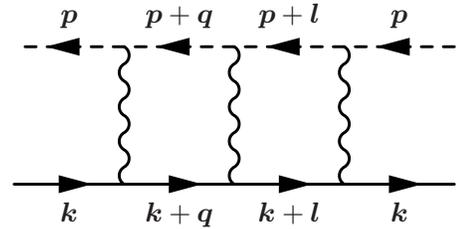}
\caption{
The particle-hole channel scattering amplitude.}
\label{fig.2Kf3}
\end{figure}

Whether the net effect is to produce a singular correction to the self energy that dominates over the bare one at low momenta and frequencies is determined by the competition between the Cooper and $2K_f$ contributions, {\em i.e} by the relative magnitude of the contribution of the two diagrams. We now argue that the Cooper channel always dominates ({\em i.e}  gives the bigger contribution) consistent with the results of the actual calculation. To see this consider both two particle scattering amplitudes when the external lines are right at the Fermi surface, and initially satisfy the `nesting' condition. In the Cooper channel this means that the total momentum of the two incoming particles is zero. In the $2K_f$ channel this means that the incoming particle-hole pair has momentum exactly $2K_f \hat{x}$. In the $2K_f$ channel exchange of a boson with momentum $q_y$
leads to a new particle-hole pair state which no longer satisfies the nesting condition (see
Fig.~\ref{fig:nesting2}). Thus after one such scattering event the particle-hole pair is less sensitive to the Amperean attraction/repulsion mediated by subsequent boson exchange.
In contrast in the Cooper channel, exchange of a boson with momentum $q_y$ preserves the nesting condition for the resulting particle-particle pair (see Fig.~\ref{fig:nesting1}). Thus they are able to continue to reap the benefits of the Amperean interaction in subsequent scattering events. This explains why the Cooper channel always dominates over the $2K_f$ channel. This difference between the kinematics of the Cooper and $2K_f$ scattering  channels is clearly present only if the Fermi surface is curved. Thus we expect that in the artificial limit where we ignore the curvature term in the fermion Greens function the two diagrams will have the same magnitude and hence will cancel. Examining the relevant integrals shows that this is exactly what happens.

Let us now discuss one important physical consequence of these results. At the nematic critical point, the singular structure of the fermion Greens
function is modified from RPA at order $1/N^2$. Specifically  the fermion Greens function satisfies the scaling form of Eqn.~\ref{frmscl} with the fermionic dynamical exponent $z = \frac{z_b}{2}$ and the exponent $\alpha = 1-\eta_f$ with $\eta_f$ given in Eqn.~\ref{etaf}. Note that $\eta_f$ is positive. This represents a modification of RPA which has $\eta_f = 0$. As pointed out in Ref. \onlinecite{critfs}, this shows up very directly in the tunneling density of states $N(\omega)$ defined through
\begin{equation}
N(\omega) = \int \frac{d^2\vec K}{(2\pi)^2} A(\vec K, \omega)
\end{equation}
where the single particle spectra function $A(\vec K, \omega)$ is defined in the usual manner:
\begin{equation}
A(\vec K, \omega) = -\frac{1}{\pi} \text{Im}G(\vec K, i\omega \rightarrow \omega+ i0^+)
\end{equation}
Here $\vec K$ is the full momentum (not linearized near the Fermi surface). Apart from being a potentially direct measure of the deviation from RPA, study of $N(\omega)$ also provides  insight into the sign of $\eta_f$ and some rationalization for why it is non-zero in the first place.  Singular contributions to $N(\omega)$ come from momenta in the vicinity of the Fermi surface.  The two dimensional momentum integral may then be separated into an angular integral over the Fermi surface and a radial integral over just the component of the momentum normal to the Fermi surface. The former just contributes an overall constant prefactor. The latter may be directly evaluated to obtain the result advertised in the introduction
\begin{equation}
N(\omega) \sim |\omega|^{\frac{\eta_f}{z}}
\end{equation}
Thus at the nematic critical point there is a power law suppression of the local single particle density of states. This suppression is of course rather natural if we remember that superconducting fluctuations are enhanced at the nematic critical point. Thus the sign of $\eta_f$ may be qualitatively understood. Further the  enhanced superconducting fluctuations make it plausible that there be some effect on the density of states unlike what happens in the RPA.

\section{Towards a phase diagram}
\label{instblty}
We now turn to the question of  what happens for general $z_b, N$. The preceding sections show that so long as $z_b - 2$ is small, the theory can be controlled for any $N$. What happens if $z_b - 2$ is not small? As discussed in the introduction for $z_b = 3$ the possibility of using large-$N$ as a control parameter has been studied in detail recently and several difficulties have been pointed out. Here we suggest an interpretation of these difficulties. Consider first the nematic critical point. The action for the order parameter fluctuations within the two patch theory was calculated to three loop order in Ref.~\onlinecite{maxsubir}. For fluctuations at zero frequency they found
\begin{equation}
\label{eqn:3loopsusc}
\int_k N \left(1 - c\sqrt{N} \right) |k_y|^2 |a(\vec k, \omega = 0)|^2
\end{equation}
The second term comes from the three loop polarizability of the fermions. The appearance of the extra factor of $\sqrt{N}$ in the loop calculation
raises concerns over the existence of a sensible large-$N$ limit. It is currently not known what the structure of higher loop terms are - for instance whether same or even higher powers of $N$ are generated by higher loop contributions. If we take the three loop answer at face value then as $c > 0$, for large enough $N$, the coefficient of $k_y^2$ becomes negative. This signals an instability towards ordering at non-zero momentum. In particular this means that the original assumption of a direct second order nematic transition is not correct, and the nematic transition will be preempted by the appearance of density wave order.

Can this conclusion be changed by higher loop diagrams? One possibility is that higher loop diagrams change the sign of the coefficient $c$. But then a different instability will likely set in. For instance consider diagrams with the same structure as those in Fig.~\ref{fig.aslamazov} but with arbitrary number of boson lines connecting the right and left moving fermions. Whatever the sign of the sum of diagrams of this sort, so long as it has a higher power of $N$ than the one-loop one, there will be an instability. If the sign is negative (as in the three loop calculation), then there is an instability where the boson likes to order at non-zero wave vector. If the sign is positive then we consider the response to an external probe field that couples with {\em opposite} sign to the two patches. A concrete example at the nematic critical point is just an external electromagnetic gauge field, {\em i.e} we consider the `diamagnetic' response to a static external magnetic field.   This changes the sign of the external vertices in diagrams like
Fig.~\ref{fig.aslamazov} without changing the sign of the internal vertices. The `diamagnetic' response to a static magnetic field then has the opposite sign from that of an ordinary metal. This signals an instability toward spontaneous flux formation, {\em i.e} the system will likely develop a state associated with spontaneous circulating currents.

The only remaining possibility is that higher loop diagrams exactly cancel the offending $\sqrt{N}$ term found in the three loop calculation. While we cannot rule this out we can provide a suggestive argument against this possibility by examining the limit of small $z_b - 2$. In this limit, the static boson polarizability may formally be written as a series
\begin{equation}
\Pi(k_y, \omega = 0) = |k_y|^{z_b - 1}\left(1 + \sum_n \frac{f_n(\lambda N; z_b)}{N^n} \right)
\end{equation}
The leading $n = 1$ term was calculated in Appendix \ref{app:3loop}. Let us assume that the functions $f_n(x;z_b)$ all have finite limits when $z_b \rightarrow 2$: \begin{equation}
\lim_{z_b \rightarrow 2} f_n(x,z_b) = F_n(x)
\end{equation}
This is explicitly seen to be true for $n = 1$, and we assume it holds for arbitrary $n$. Then successive terms in the series above are down by powers of $1/N$ for finite non-zero $\lambda N$. So for large-$N$ there is no instability. The instability potentially happens when $\lambda N$ becomes large enough that the $n = 1$ term is comparable to $1$, {\em i.e} when $\lambda N \sim N^{\frac{2}{z_b}}$. If any higher order term, say the $n$th one, is to have the same power of $N$ in the large $\lambda N$ limit (while still keeping $\epsilon$ small), then $F_n(x) \sim x^{n + \frac{z_b}{2}}$ for large  $x$. But then its coefficient has a high power of $\lambda \sim \epsilon$. Therefore any cancelation of the dangerous three loop term by higher loop diagrams cannot in general happen for arbitrary $z_b$. This makes it rather likely that there is an instability for arbitrary $z_b$ not too small.

For $z_b$ approaching $2$ comparison of the three loop term with the leading one loop term suggests that the instability happens when
$\epsilon^{\frac{2}{\epsilon}} \sim  \frac{1}{N}$. This leads to a `phase boundary' between the unstable and stable regions that comes in with infinite slope in the $\epsilon, \frac{1}{N}$ plane. Thus we propose the phase diagram shown in Fig.~\ref{favor1} or Fig.~\ref{favor2}. Through out the unstable region there is no direct nematic transition, and it is always preempted by instability toward a different broken symmetry. It is not clear whether the unstable region encompasses the all important point $N = 2, z_b = 3$, {\em i.e} whether Fig.~\ref{favor1} or Fig.~\ref{favor2} is realized. If however this point belongs to the stable region, {\em i.e} Fig.~\ref{favor1} applies and there is a direct second order nematic transition, then we have no choice but to access it from the small $\epsilon$ region (either by combining with large-$N$, or by the perturbative RG for small $N$). If Fig.~\ref{favor2} is realized on the other hand we may still hope that our expansion captures the physics at temperatures above the instability.

Similar considerations apply to the gauge field model.
There at $z_b = 3$, the gauge polarizability acquires only a positive $o(\sqrt{N})$ correction at three loop level.
However at the same order if
we consider the response to an external probe that couples with the same sign to both right and left movers,
then the sign of the three loop response is reversed.
In arguing for an instability, it is important that the bare zero-momentum susceptibility vanish,
since the result Eqn.~\ref{eqn:3loopsusc} applies in the scaling regime, which requires
$ |k_y| \ll \frac{1}{N^{3/2} \eta} $;
if the bare susceptibility were nonzero and $N$-independent, the putative instability
would occur at $k_y$ outside this regime\footnote
{
We thank Max Metlitski for
helpful correspondence on this point.
}.
A suitable choice may be a perturbation of both the volume of the system
and the chemical potential, preserving the particle density.
If the three loop calculation were the full story this would again signal an instability toward
a state which spontaneously orders at non-zero momentum and hence breaks translation symmetry.
More generally, when higher loop terms are included the situation is similar to the discussion above for the nematic model. Consequently we suggest that the gauge field model is also unstable toward a state with some broken symmetry at sufficiently large-$N$ when $z_b$ is sufficiently different from $2$. Thus once again this proposal would imply
that if the gauge model at $N= 2, z_b = 3$ is stable then we have no choice but to access it as we have done from the small $\epsilon$ region.

\section{Discussion}
\label{sec:discussion}
In this concluding section we consider the implications of our results for some specific systems, and for the general theory of non-fermi liquid metals.

An important and topical realization of the gauge model is to the theory of gapless quantum spin liquids where a gapless Fermi surface of charge neutral spin-$1/2$ fermionic spinons is coupled to a gapless fluctuating $U(1)$ gauge field. Note that in this example the spin liquid is a non-fermi liquid metal for spin transport but is an insulator for electrical transport. Such a state has been proposed\cite{lesik,leesq} to describe the intermediate temperature scale physics of  the layered organic Mott insulators $\kappa-(ET)_2Cu_2(CN)_3$ and $EtMe_3 Sb[Pd(dmit)_2]_2$.  Our controlled calculations merely confirm the correctness of several key results from RPA that are directly relevant to experiments - for instance the scaling structure of the low energy theory implies that the specific heat follows the familiar RPA result $C_v \sim T^\frac{2}{3}$ at low temperature $T$. The more important contribution of the present paper to the theory of such a spin liquid is the controlled calculation of the structure of the $2K_f$ spin correlations. Detecting these in experiments would be an interesting way to `measure' the spinon Fermi surface (see Ref. \onlinecite{norman} for
a proposal). Our results also set the stage for an analysis of phase transitions from the spinon Fermi surface state to various proximate phases with spinon pairing or other `order' that may be relevant to describing the very low temperature physics of the organics.

The model of a fermi surface coupled to a gauge field also describes algebraic charge liquid metals\cite{lee89,holonm,acl} and the related $d$-wave Bose metals\cite{dbl}. An essential difference with the particular gauge model studied in this paper is that there are two species of fermions that couple with opposite gauge charges to the same fluctuating $U(1)$ gauge field. The Amperean rules are therefore different and this will lead to some differences in the results. These can be straightforwardly handled within our expansion.
Similarly our methods are readily generalized to provide controlled expansions for various Pomeranchuk transitions other than the nematic example considered in detail in this paper.

It is interesting to contrast the quantum critical metal at these Pomeranchuk transitions with other examples of non-fermi liquid metals. One other set of examples is provided by Mott-like quantum phase transitions where an entire Fermi surface disappears continuously. Apart from continuous Mott metal-insulator transitions, these are thought to describe non-fermi liquid physics in heavy fermion metals near the onset of magnetic long range order. Ref. \onlinecite{critfs} argued that such Mott-like transitions will be characterized b the presence of a sharp critical Fermi surface but without a sharp Landau quasiparticle. A critical fermi surface is also a feature of a Pomeranchuk transition if it is second order. However we might expect that the destruction of the Landau quasiparticle is more severe at the Mott-like transitions. Indeed the explicit calculation in Ref. \onlinecite{mottcrit} for a continuous Mott transition found the exponent value $\alpha = -\eta$ (where $\eta$ is the anomalous exponent of the boson field at the $3D$ XY fixed point, and is known to be small and positive). This corresponds to a large fermion anomalous dimension $\eta_f = 1 - \alpha = 1+ \eta$. In contrast at the nematic critical point, the fermion anomalous dimension is small. Within RPA it is simply $0$ while the three loop calculation of Ref. \onlinecite{maxsubir} as well as the controlled estimate presented in this paper give non-zero but small values. The largeness of $\eta_f$ is a partial measure of the extent to which the quasiparticle is smeared (as is exemplified by the suppression of the tunneling density of states).

It is instructive to compare the non-Fermi liquids studied in this paper
with those discovered recently using holographic duality\cite{Lee:2008xf,Liu:2009dm,Cubrovic:2009ye,Faulkner:2009wj,summarypaper},
which may be understood heuristically as arising from a Fermi surface
coupled with some bath of critical fluctuations with an infinite
dynamical exponent~\cite{Faulkner:2009wj,Faulkner:2010tq,summarypaper}.
It would be interesting to find explicit field theoretical models with this feature.
In the models studied here, large $z_b > 3$ is unstable because
a $|q|^2$ term in the boson inverse propagator will always be generated by short distance fluctuations and will eventually dominate.

In summary in this paper we have developed a controlled and systematic approach to calculating the universal properties of a non-fermi liquid metal that arises when a gapless Fermi surface is coupled to a fluctuating gapless boson field. We illustrated our approach by studying spinon fermi surface spin liquids, and quantum critical metals near an electronic nematic transition in some detail. Our approach readily lends itself to the study of various closely related problems. We leave the exploration of these to the future.

\section*{Acknowledgments}
We thank Yong Baek Kim, Sung-Sik Lee, Max Metlitski and Subir Sachdev for useful discussions.
 TS was supported by
NSF Grant DMR-0705255. JM and HL were supported
 by funds provided by the U.S. Department of Energy
(D.O.E.) under cooperative research agreement DE-FG0205ER41360 and the OJI program (HL),
and by an Alfred P. Sloan fellowship (JM).

\appendix

\section{Universality and the patch construction}
\label{app:anomalies}

In this paper we have discussed a scaling theory
which focuses on the interactions
of fermions near the Fermi surface via
bosons of small frequency and small momentum.
The kinematics of these bosons allows us to restrict attention to one
patch of the Fermi surface and its antipode. In particular the universal singularities in the low energy physics is correctly
captured by breaking up the full Fermi surface into patches (and their antipodes), and studying the theory patch by patch.
In this Appendix we briefly discuss some subtle points associated with the patch construction and the subsequent treatment of the theory for any given pair of antipodal patches. As a bonus we will show that the compressibility has no singular contributions coming from the low energy scale invariant fluctuations.



Consider the following microscopic Lagrangian for the gauge field problem
\be
\label{UVmodel}
{\cal L}_{\text{UV}} = f^\dagger \(  \partial_\tau - \frac{1}{2m}\( - i \vec \nabla + \frac{1}{\sqrt{N}}\vec a \)^2 + \mu \) f ~~.
\ee
This Lagrangian is gauge-invariant,
and this forbids a mass term for the gauge field $\vec a$
in the effective action.


Next consider the low-energy description
which focuses on the modes near two antipodal patches of Fermi surface
(with normal $\hat x$, without loss of generality),
$ f \simeq f_R e^{ i k_F x } + f_L e^{ - i k_F x } $:
\bea
\nonumber
{\cal L}_{\text{patch}}
&=&
f_R^\dagger \(\partial_\tau -  i v_F \partial_x - \frac{ \partial_y^2}{2m} + \frac{v_F}{\sqrt{N}} a \) f_R
\\
&+& f_L^\dagger \(\partial_\tau + i v_F \partial_x - \frac{\partial_y^2}{2m} - \frac{v_F}{\sqrt{N}} a \) f_L
\eea
and $ v_F = k_F/m$.
Here $a$ is the $x$-component of the gauge field which couples strongly to this pair of patches. For other patches with normal $\hat{n}(\theta)$ at an angle $\theta$ to the $x$-axis, it is the component $\vec a \cdot \hat{n}(\theta)$ that will couple strongly.
The full theory is obtained by summing over all patches and adding together the `diamagnetic' term
\be
{\cal L}_{\text{dia}} \equiv \frac{a^2}{2mN} \( f^\dagger f   \)
\ee
By itself the patch action appears to respect
an `emergent' gauge symmetry which acts by
\be
\label{fakegauge}
a \to a + \partial \lambda, ~~~ f_R \to e^{ i \frac{\lambda}{\sqrt{N}}} f_R, ~~~ f_L \to e^{ i \frac{\lambda}{\sqrt{N}}} f_L
\ee
with $ \lambda = \lambda(x)$.
However, we observe that the gauge field $a$ couples in this two-patch theory to
($v_F$ times) the {\it axial} current $f_R^\dagger f_R - f_L^\dagger f_L$, which is ``anomalous",
as we now review.
To diagnose this anomaly, consider the coefficient
of $a^2$ in the effective action resulting from integrating out the fermions.
In the patch approximation, the numbers
of left- and right-moving
fermions are separately conserved.
Turning on the gauge field $a$ violates this conservation
since the gauge field couples like a chemical potential with opposite sign on the two sides.
The change in the `chiral density' of fermions
is then the density of states at the Fermi surface times
the effective chemical potential change, $\frac{v_F a}{\sqrt{N}}$.

However in the patch theory, the density of states at the Fermi surface is ill-defined.
It is apparent from its high-energy origin as a theory with a finite Fermi surface
that this description is only valid up to some maximum deviation
of the momentum from the middle of the patch. Let the patch size (which equals the cut-off for $q_y$) be denoted $\Lambda_y = K_F \Delta \theta$ where $\Delta \theta$ is the angular extent of the patch. The density of states in each patch is then
$\frac{NK_F \Delta \theta}{4\pi^2 v_F} = \frac{ Nm \Delta \theta}{4\pi^2 }$. This apparently implies that the one-loop gauge field polarizability
\be
\Pi(\vec k, \omega=0) = \vev{O(\vec k, 0) O(-\vec k,0)}_{1-loop}
\ee
depicted in Fig.~\ref{fig.1loopboson} takes a {\em non-zero value} as $\vec k \rightarrow 0$:
\be
\Pi(\vec k \rightarrow 0, \omega =0) = - \frac{m \Delta \theta}{2\pi^2 }
\ee
Note that this is the contribution from both patches to the fermion polarizability to the coefficient of $-\frac{v_F^2}{2}a^2$
in the one loop euclidean effective action for the gauge field. Naively this seems to be a problem as it violates the fake gauge invariance (\ref{fakegauge});
more problematically,
it also violates the real microscopic gauge invariance. The resolution is that the microscopic gauge invariance is obtained only when the diamagnetic term is also included in the effective action. Indeed if we sum over all patches the contribution from the polarizability is
\be
-\int_0^\pi d\theta \frac{m v_F^2}{4\pi }\left(\vec a\cdot \hat{n}\right)^2 = - \frac{K_F^2}{4\pi m} a^2
\ee
As the density satisfies $\rho_0 = \frac{N K_F^2}{4\pi}$. we see that the non-zero fermion polarizability at $\vec k \rightarrow 0, \omega = 0$ exactly cancels the diamagnetic term as required by gauge invariance.

The answer above for the fermion polarizability may be reproduced formally by considering the integral
\be
\Pi( \vec k, \omega=0) \propto \int_{p}  {\cal G}(k + p) {\cal G}(p)
\ee
Actually this integral is ill-defined at short distances and depends on the order of integration. The origins of the patch theory from the original full Fermi surface means that we need to impose a cut-off on the $x$ and $y$ momenta but not necessarily on the frequency. Doing the frequency integral by contour integration we reproduce (in the limit $\vec k \rightarrow 0$) the non-zero constant obtained above through a physical argument.
Note that imposing a hard cut-off on $p_x$  violates the fake gauge invariance of
Eqn.~\ref{fakegauge}. Thus a careful formulation of the patch theory that is faithful to its microscopic origins requires imposing a cut-off on the fermion momenta, and taking the frequency cut-off to infinity first before sending the momenta cut-offs to infinity to define the scaling limit.

However as in any scaling theory we expect that universal low energy singularities are actually insensitive to how the theory is regularized at short distances ({\em i.e} independent of the ``short distance completion" of the theory). Thus for the purpose of calculating the universal singularities we are free to choose any regularization of the patch theory. A convenient choice (and the one used in this paper) is to define the scaling theory by sending the momenta cut-offs to infinity first, and then the frequency cut-offs. Then the polarization integral may be done by first integrating over $p_x$. The result then vanishes as both poles of $p_x$ lie on the same side in the complex plane. Thus in this regularization of the patch theory there is no need to include a diamagnetic term to maintain gauge invariance. Indeed with this regularization the fake gauge invariance is no longer fake and is a real property of the universal scaling theory.

Though this choice for defining the scaling theory will reproduce all the universal singularities there is no guarantee that it will correctly reproduce non-universal ones as it is not faithful to the original microscopic action with the full Fermi surface. A good example to illustrate this is the fermion compressibility, {\em i.e} the response in the density of the system to a change of chemical potential. Within the patch construction the chemical potential couples to $f^\dagger_R f_R + f^\dagger_L f_L$. This is identical to the coupling of the nematic order parameter. In the single patch theory, the chemical potential term is thus identical to the gauge coupling term. The contribution to the compressibility is thus given by the same $\Pi(\vec k \rightarrow 0, \omega = 0)$ discussed above. With the regularization actually employed in this paper (which dictates us to do the $p_x$ integral first) this is zero as is indeed required by the gauge invariance of
Eqn.~\ref{fakegauge}. In the other regularization (dictated by the original microscopic situation) we do the frequency integral first and get a non-zero answer.

What is the interpretation of the zero answer that doing the $p_x$ integral first produces? Clearly it means that there is no {\em singular} contribution to the compressibility in the low energy theory. The non-zero answer found in the other regularization must be viewed as a smooth non-singular background that is not correctly described by the scale invariant patch theory. The validity of Eqns. \ref{fakegauge} in the scaling limit employed in the paper thus ensures that singular contributions to the compressibility vanish to all orders of perturbation theory in the one patch theory. Actually the same result is also true in the two patch theory.  To see this note that with the short distance completion we have chosen the two patch action in the presence of an external potential that couples to $f_R^\dagger f_R + f_L^\dagger f_L$ enjoys a low energy gauge invariance similar to Eqn.~\ref{fakegauge} but where we rotate the phases of right and left movers with opposite phases $\pm \frac{\lambda}{\sqrt{N}}$. This gauge invariance ensures that the contribution to the fermion compressibility vanishes even in the two patch theory. For the original microscopic model this then implies that there is no singular contribution to the background non-zero compressibility.

In Appendix \ref{app:orderofintegration},
we explicitly check that short-distance
sensitivity (similar to what happened in the polarizability integral) does not arise in some of the universal quantities of interest,
and the answers are insensitive to the order of integration.

\section{Example diagram - single patch}
\label{app:1patchdia}
\begin{figure}[ht]
\centering
\includegraphics[width=42mm]{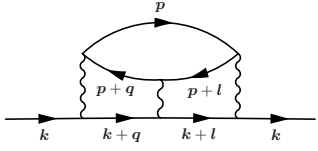}
\caption{Three-loop fermion self-energy containing fermions from a single patch.}
\label{fig.3loop}
\end{figure}

Consider the 3-loop self-energy contribution shown in Fig.~\ref{fig.3loop} with the external Fermion on the Fermi-surface, i.e. $\epsilon^\pm_k\equiv  \pm k_x+k_y^2=0$ (``$+$" and ``$-$" for the right and left patch, respectively) for $z_b\rightarrow 2$.
\normalsize
\begin{eqnarray}
\Sigma^{(3)}=&&\frac{1}{N^2} \int d\bar{\vect{l}}d\bar{\vect{q}}d\bar{\vect{p}} D(\vect{q})D(\vect{q}-\vect{l})D(\vect{l}){\cal G}_{R/L}(\vect{p})\\
&&\times {\cal G}_{R/L}(\vect{k}+\vect{q}){\cal G}_{R/L}(\vect{k}+\vect{l}){\cal G}_{R/L}(\vect{p}+\vect{q}){\cal G}_{R/L}(\vect{p}+\vect{l}),\nonumber
\end{eqnarray}
where
\begin{align}{\cal G}_{R/L}^{-1}(\vect{k})&=-\frac{\i}{\lambda N} \text{sgn}(\omega_k)|\omega_k|^{2/z_b}+\epsilon_k^\pm
\end{align}
and $d\bar{\vect{l}}\equiv \frac{d\omega_ldl_xdl_y}{(2\pi)^3}$.

Begin by integrating out $\vect{p}=(\omega_p,\vec{p})$:
\begin{eqnarray}
&&\int d\bar{\vect{p}}{\cal G}_{R/L}(\vect{p}+\vect{q}){\cal G}_{R/L}(\vect{p}){\cal G}_{R/L}\vect{p}+\vect{l})\nonumber\\
&&=\frac{\omega_l f(l_y,q_y,\omega_l,\omega_q)-\omega_q f(q_y,l_y,\omega_q,\omega_l)}{q_y(\tilde\Sigma(\omega_l)-\epsilon_l^\pm)-l_y(\tilde\Sigma(\omega_q)-\epsilon_q^\pm)},
\end{eqnarray}
where $f$ is a non-singular function independent of $z_b$ and $N$. The $l_x$ and $q_x$ integrations are straightforward and we find
\small
\begin{align}
\Sigma^{(3)}= &\frac{\i \text{sgn}(\omega_k)|\omega_k|^{2/z_b} \lambda N}{N^2}\int dq_y ql_y d\omega_q d\omega_l \text{sgn}(q_y-l_y)\nonumber\\
& \ \ \ \times \frac{\omega_l \tilde f_{q,l}-\omega_q \tilde  f_{l,q}}{\gamma|\omega_q-\omega_l|+|q_y-l_y|^{2}} D(\vect{q})D(\vect{l}).
\end{align}
\normalsize
One still needs to verify that the remaining integral is finite. In Ref.\onlinecite{ssl} it has been shown that all planar diagrams in the single-patch theory are UV-finite, so it remains to verify IR-finiteness. For $\vect{l}$ finite and $\vect{q} \rightarrow 0$ or $\vect{q}\rightarrow \vect{l}$ this is clear. For $\vect{q},\vect{l}\rightarrow 0$ the integrand diverges as $q_y^{-2}$ which is  cancelled by the remaining integrations. Thus the 3-loop self energy is indeed of order $N^{-2}$ as long as $\lambda N$ is of order unity.

\section{Self energies at three loops}
\label{app:3loop}
Ref.~\onlinecite{maxsubir} identified the important diagrams that contribute to the boson and fermion self energies at three loop level, and evaluated them in appropriate limits when $z_b = 3$. Here we briefly sketch the modifications for general $z_b$ close to $2$.
Consider the diagrams in Fig.~\ref{fig.aslamazov} that determine the boson self energy at three loops.
For $z_b \approx 2$ we find
\begin{align}\Pi&=|k_y|^{z_b-1}\frac{2 \lambda N \gamma }{ N \pi}\int_0^{1} d s\int_0^\infty d t \frac{(1-s)^{3} t}{t+1}\frac{s}{ t (1-s)^{2}+s^{2}}\nonumber \\
&\times  \frac{\(s\gamma\lambda N\)^2}{(1-s)^2 t^2+\(s\gamma\lambda N\)^2}.
\end{align}
In the physical limit $\gamma \lambda =\frac{\pi}{2}$, $N=2$ we evaluate the integral numerically to find
\begin{align}\Pi&=0.106|k_y|^{z_b-1}.
\end{align}

Now consider the diagrams  for the  three loop fermion self-energy contributions depicted in
Fig.~\ref{fig.3looplr}.

\begin{figure}[ht]\centering
\includegraphics[width=80mm]{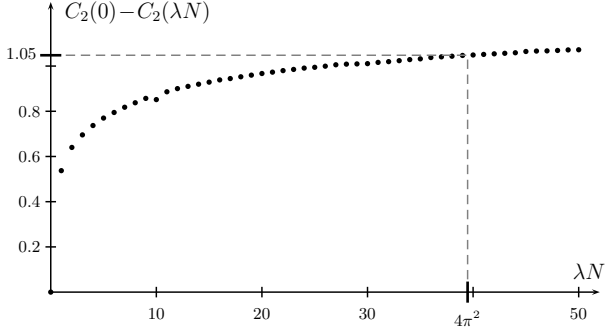}
\caption{Numerical integral determining the fermion anomalous dimension $\eta_f$.}
\label{fig.plot}
\end{figure}

They are given by
\begin{align}
\Sigma^{(3)}_{1,2}&=-\frac{1}{N^2} \int_{\bar\omega_l,\vec{\bar{l}}}\int_{\bar\omega_q,\vec{\bar{q}}}\int_{\bar\omega_p,\vec{\bar{p}}}D(\vect{q})D(\vect{p})D(\vect{q}-\vect{p})\\
& \times {\cal G}_{R}(\vect{k}\pm\vect{p}){\cal G}_{R}(\vect{k}\pm\vect{q}){\cal G}_{L}(\vect{l}){\cal G}_{L}(\vect{l}+\vect{q}){\cal G}_{L}(\vect{l}+\vect{p}).\nonumber
\end{align}
The integrals may be evaluated following Ref.~\onlinecite{maxsubir}, the results are, to leading order in $\epsilon_k$:
\begin{align}
\Sigma^{(3)}_1(\epsilon_k)&= \frac{\lambda^2}{ 12\pi^4 } C_{z_b}(0) \epsilon_k \ln \frac{\Lambda}{\epsilon_k}\\
\Sigma^{(3)}_2(\epsilon_k)&=\Sigma^{(3)}_2(0)- \frac{\lambda^2}{12\pi^4 } C_{z_b}(\lambda N) \epsilon_k \ln \frac{\Lambda}{\epsilon_k^{2/z_b}},
\end{align}
where
\begin{widetext}
\small
\begin{align}
C_{z_b}(\lambda N)=&\frac{96 }{(4\pi)^{\frac{4}{z_b}}}\int\limits_1^\infty dx \int\limits_0^\infty dy \int\limits_0^\infty ds \int\limits_s^\infty dt \frac{t s (s-t)^2}{(x+s^{z_b})(y+t^{z_b})(x+y+(t-s)^{z_b})}\\
&\times\frac{\big(t\big((x-1)^{\frac{2}{z_b}}+x^{\frac{2}{z_b}}+1\big)    +s \big((y+1)^{\frac{2}{z_b}}+y^{\frac{2}{z_b}}-1\big) \big)^2-  N^2 \lambda^2(4\pi)^{-\frac{2}{z_b}} s^2t^2(s-t)^2}{\big(\big(t\big((x-1)^{\frac{2}{z_b}}+x^{\frac{2}{z_b}}+1\big)    +s \big((y+1)^{\frac{2}{z_b}}+y^{\frac{2}{z_b}}-1\big) \big)^2+  N^2 \lambda^2(4\pi)^{-\frac{2}{z_b}} s^2t^2(s-t)^2\big)^2}\nonumber\\
+&\frac{48}{(4\pi)^{\frac{4}{z_b}}}\int\limits_1^\infty dx \int\limits_1^\infty dy \int\limits_0^\infty ds \int\limits_0^\infty dt \frac{t s (s+t)^2}{(x+s^{z_b})(y+t^{z_b})(|x-y|+(t+s)^{z_b})}\nonumber\\
&\times\frac{\big(t\big((x-1)^{\frac{2}{z_b}}+x^{\frac{2}{z_b}}+1\big)    +s \big((y-1)^{\frac{2}{z_b}}+y^{\frac{2}{z_b}}+1\big) \big)^2-  N^2 \lambda^2(4\pi)^{-\frac{2}{z_b}} s^2t^2(s+t)^2}{\big(\big(t\big((x-1)^{\frac{2}{z_b}}+x^{\frac{2}{z_b}}+1\big)    +s \big((y-1)^{\frac{2}{z_b}}+y^{\frac{2}{z_b}}+1\big) \big)^2+  N^2 \lambda^2(4\pi)^{-\frac{2}{z_b}} s^2t^2(s+t)^2\big)^2}\nonumber.
.\end{align}
\normalsize
\end{widetext}
The function $J(\lambda N)$ which appears in Eq. (\ref{eqn:jequation}) is the sum of the singular contributions at $z_b=2$, i.e.
\begin{align}
J(\lambda N)=\(\frac{\lambda N}{4\pi^2}\)^2\(C_{2}(0)-C_{2}(\lambda N)\)
\end{align}
(note that the prefactor is unity at $z_b=3$, $N=2$). A numerical estimate of this function is shown in Fig.~\ref{fig.plot}.

\section{Vertex integral}
\label{app:vertex}

\begin{figure}[h]
\centering
\includegraphics[width=60mm]{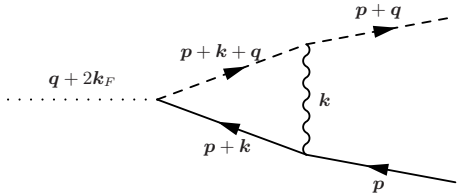}
\caption{One loop correction to the $2k_\text{F}$-vertex. The dotted line denotes a $2k_\text{F}$ density fluctuation.}
\label{fig.vertex}
\end{figure}
We want to evaluate the integrals shown in Fig.~\ref{fig.vertex}, where all external momenta and fequencies are put to zero. The computation is straightforward:
\begin{eqnarray}
\delta u &=&-\frac{1}{N} \int_{\vec k, \omega}\frac{ D(k_y, \omega)}{\i\text{sgn}(\omega)\frac{|\omega|^{\frac{2}{z_b}}}{\lambda N}-\epsilon_k^+}\frac{1}{\i\text{sgn}(\omega)\frac{|\omega|^{\frac{2}{z_b}}}{\lambda N}-\epsilon_k^-} \nonumber\\
&=& -\frac{1}{2 N} \int_{k_y, \omega}\frac{\frac{1}{\lambda N}|\omega|^{\frac{2}{z_b}}}{\frac{1}{\lambda^2 N^2}|\omega|^{\frac{4}{z_b}}+k_y^4}\frac{|k_y|}{\gamma |\omega|+|k_y|^{z_b}}\nonumber\\
&=&- \frac{4}{2 N (2\pi)^2}\int_{\Lambda/\sqrt{b}}^{\Lambda} dk_y\frac{1}{k_y} \int_0^\infty d t\frac{\frac{1}{\lambda N}t^{\frac{2}{z_b}}}{\frac{1}{\lambda^2 N^2}t^{\frac{4}{z_b}}+1}\frac{1}{\gamma t+1}\nonumber\\
&=& \frac{1}{4\pi^2 N}\ln b \int_0^\infty d t\frac{\lambda N t^{\frac{2}{z_b}}}{t^{\frac{4}{z_b}}+\lambda^2 N^2}\frac{1}{\gamma t+1}.
\end{eqnarray}
The corresponding expression at the perturbative (small $\epsilon$) fixed point is given by
\begin{eqnarray}
\delta u &=& \frac{1}{4\pi^2 N}\ln b \int_0^\infty d t\frac{ t^{1-\frac{\epsilon}{2}}}{t^{2-\epsilon}+1}\frac{1}{ t+\frac{2}{\pi \epsilon N}}.
\end{eqnarray}
The integral remains finite if we take the limit $\epsilon \rightarrow 0$ in the first term in the integrand, so we can evaluate it analytically. We obtain to leading order in $\epsilon$
\begin{eqnarray}
\delta u &=& \frac{ \epsilon }{8 \pi}  \ln \frac{2}{\pi \epsilon N}\ln b
\end{eqnarray}

\section{Order of integration}
\label{app:orderofintegration}
Consider  as a concrete example the three-loop self energy depicted in Fig.~\ref{fig.3looplr}.
We want to keep a cut-off on they $y$-components of momenta and verify that we can interchange the order in which we perform the integration over frequencies and momentum $x$-components.  As the only possible divergencies arise from the UV, we can set all $y$-components as well as external momenta and frequencies to zero (the $y$ components in the numerators of the gauge propagators are set to unity). Recall that any two integrals are independent of the order of integration, if the double integral is absolutely convergent. So consider
\begin{align}
&\int_{l_x,p_x,q_x,\omega_l,\omega_p,\omega_q}\hspace{-1cm}| D_{\omega_q}D_{\omega_q-\omega_l}D_{\omega_l}{\cal G}_{\vect{p}} {\cal G}_{\vect{q}}{\cal G}_{\vect{l}}{\cal G}_{\vect{p}+\vect{q}}{\cal G}_{\vect{p}+\vect{l}}|\\
 \sim &\int d\omega_l \frac{|\omega_l|^{6/z_b}}{|\omega_l^{10/z_b}|}\int d\omega_p \frac{|\omega_p|^{1+6/z_b}}{|\omega_p^{2+10/z_b}|}\int d\omega_q \frac{|\omega_q|^{2+6/z_b}}{|\omega_q^{3+10/z_b}|}\nonumber\\
& \ \ \ \times\int dp_x\frac{1}{|p_x|^2}\int dq_x\frac{1}{|q_x|^2}\int dl_x\frac{|l_x|^2}{|l_x|^5}.
\end{align}
Since each of these integrals is individually convergent, the original integrations can be performed in any order. This analysis applies to most of the diagrams shown here and is a result of the Landau-damping form of the gauge propagator, thus it does not extend to diagrams calculated at the small $\epsilon$ pertubative fixed point. In our proposed expansion $N\rightarrow \infty$, $\lambda N =o(1)$, of the diagrams shown in this paper only the one-loop boson self-energy (Fig.~\ref{fig.1loopboson}) is sensitive to the order of integration.

\end{document}